\documentclass[final,onefignum,onetabnum]{siamart171218}

\usepackage{amssymb}
\usepackage[normalem]{ulem}
\newcommand\redout{\bgroup\markoverwith{\textcolor{red}{\rule[.5ex]{2pt}{.8pt}}}\ULon}
\newcommand\blueout{\bgroup\markoverwith{\textcolor{blue}{\rule[.5ex]{2pt}{.8pt}}}\ULon}
\newcommand\magentaout{\bgroup\markoverwith{\textcolor{magenta}{\rule[.5ex]{2pt}{.8pt}}}\ULon}
\usepackage{lipsum}
\usepackage{amsfonts}
\usepackage{graphicx}
\usepackage{epstopdf}
\usepackage{algorithmic}
\ifpdf
  \DeclareGraphicsExtensions{.eps,.pdf,.png,.jpg}
\else
  \DeclareGraphicsExtensions{.eps}
\fi

\newsiamremark{remark}{Remark}
\newsiamremark{hypothesis}{Hypothesis}
\crefname{hypothesis}{Hypothesis}{Hypotheses}
\newsiamthm{claim}{Claim}

\headers{fast marching method with superior performance}{J. Yang}

\title{An easily-implemented, block-based fast marching method with superior sequential and parallel performance\thanks{Submitted to the editors September 10, 2018.}}
\author{Jianming Yang\thanks{Fidesi Solutions LLC, P.O. Box 734, Iowa City, IA 52244, USA 
  (\email{jmyang@fidesisolutions.com}).}}
\usepackage{amsopn}

\ifpdf
\hypersetup{
  pdftitle={An easily-implemented, block-based fast marching method with superior sequential and parallel performance},
  pdfauthor={J. Yang}
}
\fi

\begin{document}

\maketitle

\begin{abstract}
The fast marching method is well-known for its worst-case optimal computational complexity in solving the Eikonal equation, and has been employed in numerous scientific and engineering fields. However, it has barely benefited from fast-advancing multi-core and many-core architectures, due to the challenges presented by its apparent sequential nature. In this paper, we present a straightforward block-based approach for a highly scalable parallelization of the fast marching method on shard-memory computers. Central to our new algorithm is a simplified restarted narrow band strategy, with which the global bound for terminating the front marching is replaced by an incremental one, increasing by a given stride in each restart. It greatly reduces load imbalance among blocks through a synchronized exchanging step after the marching step. Furthermore, simple activation mechanisms are introduced to skip blocks not involved in either step. Thus the new algorithm mainly consists of two loops, each for performing one step on a group of blocks. Notably, it does not contain any data race conditions at all, ideal for a direct loop level parallelization using multiple threads. A systematic performance study is carried out for several point source problems with various speed functions on four grids with up to $1024^3$ points using two different computers. Substantial parallel speedups, e.g.,  up to 3--5 times the number of cores on a 16-core/32-thread computer and up to two orders of magnitude on a 64-core/256-thread computer, are demonstrated with all cases. As an extra bonus, our new algorithm also gives improved sequential performance in most scenarios. Detailed pseudo-codes are provided to illustrate the modification procedure from the single-block sequential algorithm to the multi-block one in a step-by-step manner.
\end{abstract}

\begin{keywords}
Eikonal equation, Fast marching method, Narrow band approach, Domain decomposition, Parallel algorithm, Shared-memory parallelization
\end{keywords}

\begin{AMS}
35F30, 35L60, 49L25, 65N06, 65N22, 65K05
\end{AMS}

\section{Introduction} \label{sec:intro}

The fast marching method  \cite{Sethian96,Sethian99a,Sethian99b} solves the stationary boundary value problem defined by the Eikonal equation:
\begin{equation}
\begin{array}{cc}
 | \nabla \psi (\mathbf{x}) | F (\mathbf{x}) = 1, & \mathbf{x} \in \varOmega \, \backslash \, \varGamma, \\
\psi (\mathbf{x}) = 0, & \mathbf{x} \in \varGamma \subset \varOmega,
\end{array}
\label{eq:eikonal}
\end{equation}
where $\varOmega$ is a domain in $\mathcal{R}^n$, $\varGamma$ is the initial boundary, and  $F (\mathbf{x})$  is a positive speed function, with which the boundary information propagates away along characteristics in the domain. This equation has been used in  a variety of applications, such as computational geometry, computer vision, optimal control, computational fluid dynamics, materials science, etc. The fast marching method is a non-iterative algorithm that closely resembles Dijkstra's method \cite{Dijkstra59} for finding the shortest path on a network. It was first proposed by Tsitsiklis \cite{Tsitsiklis95} using an optimal control approach. Sethian \cite{Sethian96} derived the algorithm based on an upwind difference scheme and introduced the name ``fast marching'', which has since become a very popular method for solving the Eikonal equation. 

The success of the fast marching method stems from its simplicity and efficiency. The discretization of the Eikonal equation using upwind difference schemes is straightforward. At a given grid point, the finite difference stencil contains only upwind neighbors and the causality of the equation is strictly followed. In terms of data structure, it only involves a heap priority queue, which is used to march the solution in a rigorous increasing order. In the fast marching method, the number of times that a point is visited is minimized and no iterations are involved in the one-pass solution updating process. For a case with $N$ grid points, the fast marching method has a worst-case optimal algorithm complexity of $O(N \log N)$, as the run-time complexity of reordering of a heap of length $n$ is $O(\log n)$.

Several studies have been reported to improve the $O(N \log N)$ run-time complexity to obtain $O(N)$ complexities for solving the Eikonal equation in a single pass. In particular, Tsitsiklis proposed a $O(N)$ approach using a bucket data structure along with his $O(N \log N)$ Dijkstra-like method \cite{Tsitsiklis95}. Since insertion and deletion of a node in a bucket structure takes only $O(1)$ computations instead of $O(\log n)$ in the binary heap implementation of a priority queue, the complexity of the whole algorithm was reduced to $O(N)$. A shared-memory parallel implementation was also provided. In \cite{YatzivBS06}, an $O(N)$ implementation of the fast marching method was developed with the help of untidy priority queue that does not distinguish the priorities of the points within the same bucket. However, the quantization of the priorities introduces an extra error of the same order of magnitude of the discretization scheme. The $O(N)$ methods in \cite{Tsitsiklis95} and \cite{YatzivBS06} are single-pass approaches like the original fast marching method. In contrast, the group marching method developed in \cite{Kim01} obtained an $O(N)$ complexity without the use of sorting data structures. In this method, essentially, a group of points in a narrow band from the vicinity of the marching front are set to be valid at the same time through two passes of solution updating. However, these $O(N)$ algorithms seem to have received less attention than other fully iterative algorithms of $O(N)$ complexity, such as the fast sweeping methods \cite{Zhao05} and the fast iterative methods \cite{JeongW08}. 

A pronounced feature of iterative methods is that they are highly susceptible to different parallelization strategies, either coarse grained \cite{Zhao07,Gillberg2014,ChaconV2015} or fine grained \cite{JeongW08,WeberDBBK08,DetrixheGM2013}. On the contrary, the fast marching method has no similar straightforward parallelism as found in these iterative methods. Actually, it was long deemed inherently sequential due to the use of a heap priority queue, with which there is only one single point in the whole domain to be accepted as valid at a time. A few attempts to acheive parallel implementations using domain decomposition techniques were reported in the literature \cite{Herrmann03,BreussCGV2011}, but generally with limited success until recently. In \cite{yang2017highly}, a highly scalable massively parallel algorithm of the fast marching method was established through a rigorous systematical test.  In this method, essentially, a slightly modified sequential fast marching algorithm was run on each process for one subdomain. A novel restarted narrow band approach was introduced to coordinate the synchronization of front propagation among subdomains, such that the the frequency of inter-process communications and the number of points to be refreshed with incoming front characteristics could be balanced to minimize the total cost. Several other means, such as extended data structures for interface problems, and augmented tags for point status, were also introduced to streamline the algorithm. This method was based on an overlapping domain decomposition technique for distributed-memory parallelization and developed for large-scale practical applications of today using billions of grid points on hundreds of thousands of processes. However, it is still hindered by the load imbalance inherent to the solution process of the Eikonal equation in a stationary domain decomposition setting. That is, a process might stay idle until the advancing front engages the subdomain assigned to it because of the fixed one-to-one process-subdomain relationship. In addition, the MPI-based parallel algorithm could not take full advantage of the shared-memory architecture prevalent in multi-core and/or many-core processing.

In this paper, we address these issues with a fresh view on the sequential and parallel fast marching algorithms mentioned above, and propose a block-based approach with superior sequential and parallel peformance. This approach follows the original sequential algorithm nearly exactly for a single block, obtains mostly much improved performance immediately with multiple blocks, and almost alway realizes a parallel efficiency way above unity with multiple threads working on these blocks. The restarted narrow band strategy \cite{yang2017highly} is still central to the whole process, as it is fully consistent with the narrow band fast marching method. Nonetheless, a non-overlapping domain decomposition technique is adopted. Compared with the overlapping approach used in \cite{yang2017highly}, the non-overlapping approach complies better with the sequential algorithm and facilitates completely local computation in each subdomain. Thus the whole algorithm does not involve any race conditions, which usually require the introduction of lock synchronization, a common performance bottleneck in shared-memory parallelization. Moreover, simple block activation-deactivation mechanisms are designed to determine if a block should be included in the narrow band marching step or the  ghost cell data exchanging step. Therefore, the load imbalance issues prominent in a stationary domain decomposition setting are substantially mitigated. Starting with a standard sequential fast marching algorithm, we introduce small modifications on top of it in a  step-by-step manner without altering any algorithm essentials. Together with systematic tests on different platforms, an illustrative approach is undertaken to provide a dependable benchmark parallel algorithm for the sequential fast marching method.

\section{Numerical Methods}\label{sec:numerical}

In this part, a color code is used in the pseudo-codes to illustrate the incremental modifications introduced in each Subsection: 1) black, for the original sequential fast marching method in Sec. \ref{sec:sfmm}; 2) red, for augmented status tags and the unified heap in Sec. \ref{sec:uniheap}; 3) green, for the non-overlapping domain decomposition technique in Sec. \ref{sec:nbfmm}; 4) blue, for the restarted narrow band strategy in Sec. \ref{sec:restarted}; 5) cyan, for the block activation mechanism for the marching step; and 6) magenta, for the block activation mechanism for the exchanging step. The shared-memory parallelization using OpenMP threads is not given here, because it is fairly straightforward based on the current pseudo-codes, and particular OpenMP directives depend on the programming language used for the implementation. Also, for brevity, interface problems are not discussed here, but would be treated exactly the same as presented in \cite{yang2017highly}.

\subsection{Sequential fast marching method}\label{sec:sfmm}

To simplify the discussion, we use a uniform grid with a cell size of $\Delta h$ in each direction to discretize the physical domain $\varOmega$, although the methodology is by no means restricted to such a uniform grid. As shown in Fig. \ref{fig:domain}(a), the unknown variable, $\psi$, is defined at the cell center (i.e., the grid point) and the grey-shaded zones are filled with ghost cells. For conenience, the ghost cell zones are collectedly named $\varTheta$ and $\varXi = \varOmega + \varTheta $ as the combination of the discretized physical domain and ghost cell zones. The following quadratic equation for $\psi_{i,j,k}$ can be obtained using the Godunov-type upwind finite difference scheme \cite{RouyT92} to approximate Eq. (\ref{eq:eikonal}):
\begin{equation}
 \begin{array}{cc}
 &    \left( \max (D^{-x}_{i,j,k} \psi, -D^{+x}_{i,j,k} \psi, 0) \right ) ^2\\
 + &  \left( \max (D^{-y}_{i,j,k} \psi, -D^{+y}_{i,j,k} \psi, 0) \right ) ^2\\
 + &  \left( \max (D^{-z}_{i,j,k} \psi, -D^{+z}_{i,j,k} \psi, 0) \right ) ^2
\end{array}
  = \dfrac{1}{F ^2 _{i,j,k}},
\label{eq:fds}
\end{equation}
where the operators $D^{-x}_{i,j,k}$ and $D^{+x}_{i,j,k}$ define the backward and forward first-order difference approximations to the spatial derivative $\partial \psi / \partial x$, respectively:
\begin{equation}
D^{-x}_{i,j,k} \psi = \dfrac{\psi_{i,j,k} - \psi_{i-1,j,k}}{\Delta h}, \quad D^{+x}_{i,j,k} \psi = \dfrac{\psi_{i+1,j,k} - \psi_{i,j,k}}{\Delta h}.
\label{eq:fos}
\end{equation}
The $y$ and $z$ directions are treated similarly.

From Eq. (\ref{eq:fds}) it is apparent that the solution of $\psi_{i,j,k}$ only relies on the neighboring points of smaller value, or upwind points. The fast marching method takes advantage of this fact by following a systematical way of solving Eq. (\ref{eq:fds}) in the computational domain starting from the point with the smallest value. To establish the order of updating, the grid points are divided into three categories:  i) $\texttt{KNOWN}$, which contains points that are part of the boundary condition or considered to have final solutions; ii) $\texttt{BAND}$, which contains points that have $\texttt{KNOWN}$ neighbor(s), thus their solutions can be updated by solving Eq. (\ref{eq:fds}); and iii) $\texttt{FAR}$, which contains points that do not have any $\texttt{KNOWN}$ neighbors yet, thus it is unnecessary to solve Eq. (\ref{eq:fds}) on them.  For a new $\texttt{KNOWN}$ point, the solutions of its $\texttt{BAND}$ and $\texttt{FAR}$ neighbors can be updated through Eq. (\ref{eq:fds}). Then the point that has the smallest value in the $\texttt{BAND}$ set is removed from the $\texttt{BAND}$ set to become a new $\texttt{KNOWN}$ point, which clearly has the largest solution in all $\texttt{KNOWN}$ points. This loop continues until all points are $\texttt{KNOWN}$ or the solution of the last $\texttt{KNOWN}$ point reaches a pre-set threshold. In order to maintain a strict order of increasing values in this process, a binary heap structure is usually adopted to sort points in the $\texttt{BAND}$ set \cite{Sethian99b}. Several standard heap operations are involved, e.g., $\texttt{Insert\_Heap}$ adds an element into the heap, $\texttt{Locate\_Min}$ gives the index of the top element on the grid, $\texttt{Remove\_Min}$ deletes the top element from the heap, $\texttt{Up\_Heap}$ moves an element up in the heap, etc. 

\begin{algorithm}
\algsetup{indent=1em}
\caption{Heap initialization:\newline $\textsc{Initialize\_Heap}{\color{green}{(ib)}}$.}
\label{alg:init_heap}
\begin{algorithmic}[1]
  \STATE $\psi ^{\color{green}{ib}} \leftarrow +\infty$
  \STATE $G ^{\color{green}{ib}} \leftarrow  \texttt{FAR}$
  \STATE $\texttt{size} (\mathfrak{H} ^{\color{green}{ib}} ) \leftarrow 0$
  {\color{green}{\STATE {$\mathfrak{A} ^{ib}    \leftarrow  \texttt{FALSE}$}}}
  \FORALL{$(i,j,k)\in \varXi ^{\color{green}{ib}} $ }
    \IF{ $(i,j,k)$ is on or adjacent to $\Gamma$} 
      \STATE $\psi ^{\color{green}{ib}} _{i,j,k}\leftarrow  \psi_0$
      \STATE $G ^{\color{green}{ib}} _{i,j,k} \leftarrow  \texttt{KNOWN\color{red}{\_FIX}}$ 
      \STATE {\redout{$\textsc{Update\_Neighbors}(i,j,k {\color{green}{,ib}})$}} \color{red}{$\textsc{Insert\_Heap}(i,j,k,\mathfrak{H}^{\color{green}{ib}})$}
          {\color{green}{\STATE {$\mathfrak{A} ^{ib}  \leftarrow  \texttt{TRUE}$}}}
    \ENDIF
  \ENDFOR
\end{algorithmic}
\end{algorithm}

Therefore, this first step of the narrow band fast marching algorithm is the initialization of the heap structure, as shown in Algorithm \ref{alg:init_heap}. In this work, a status tag $G$ is used to label the category that a point $(i,j,k)$ belongs to. All points in $\varXi$ are initialized with a function value $ + \infty$ and a status tag $\texttt{FAR}$. The initial heap size is set to be zero. Then, the grid points on or immediately adjacent to the boundary are specified with initial function values $\psi_0$ and marked as $\texttt{KNOWN}$. The neighbors of a $\texttt{KNOWN}$ point can be updated as shown in Algorithm \ref{alg:update}. Here only the neighboring points within $\varOmega$ that are not $\texttt{KNOWN}$ are checked by solving the quadratic equation using Algorithm \ref{alg:quadratic}. If the solution satisfies the causality condition and is smaller than the current value, it can be accepted. Correspondingly this point is tagged as $\texttt{BAND}$ and inserted to the heap or moved up in the heap if it was added earlier. After the initialization is completed, the front can be marched in a loop as shown in Algorithm \ref{alg:march_new}. Obviously, if the heap is empty or if the top element in the heap has a value larger than the preset narrow band bound, $\texttt{width}_{\text{band}}$, the marching of front should be terminated. Otherwise, the top element is removed from the heap and tagged as $\texttt{KNOWN}$. Then its neighbors can be updated as discussed above. The operations within the loop repeat until one of the exit conditions is met. For more details of the sequential narrow band fast marching method, the reader is referred to \cite{Sethian96,Sethian99a,Sethian99b} and the references therein.

\begin{algorithm}
\algsetup{indent=1em}
\caption{Update neighbors of a point just removed from the heap:\newline $\textsc{Update\_Neighbors}(i,j,k{\color{green}{,ib}})$.}
\label{alg:update}
\begin{algorithmic}[1]
\FORALL{$(l,m,n)$ such that $\left( |l-i|+|m-j|+|n-k|\right) = 1$}
  \IF{$(l,m,n) \in \Omega ^{\color{green}{ib}} $}
    \IF{$G ^{\color{green}{ib}} _{l,m,n} \notin \texttt{KNOWN{\color{green}{\_FIX}}}$ $\color{green}{\mathbf{and}\; \psi ^{\color{green}{ib}} _{l,m,n}   > \psi ^{\color{green}{ib}} _{i,j,k} } $}
      \STATE $\psi_{\text{temp}} \leftarrow \textsc{Solve\_Quadratic}(l,m,n{\color{green}{,ib}})$
        \IF{$\psi_{\text{temp}} < \psi ^{\color{green}{ib}} _{l,m,n}  $}
          \STATE $\psi ^{\color{green}{ib}} _{l,m,n}   \leftarrow \psi_{\text{temp}}$
          \STATE $G ^{\color{green}{ib}} _{l,m,n}   \leftarrow \texttt{BAND}$
          \IF{$(l,m,n) \not\in \mathfrak{H} ^{\color{green}{ib}} $} 
            \STATE $\textsc{Insert\_Heap}(l,m,n,\mathfrak{H} ^{\color{green}{ib}})$ 
          \ELSE   
            \STATE $\textsc{Up\_Heap}(l,m,n,\mathfrak{H} ^{\color{green}{ib}})$ 
        \ENDIF
            \color{magenta}{\IF{$(l,m,n) \in \Pi ^{ib} $} 
            \STATE {$\Upsilon ^{ib} \leftarrow \Upsilon ^{ib} \cup \{(l,m,n)\}$}
             \ENDIF}
      \ENDIF
    \ENDIF
  \ENDIF
\ENDFOR
\end{algorithmic}
\end{algorithm}

\subsection{Augmented status tags and unified heap}\label{sec:uniheap}

In \cite{yang2017highly}, augmented status tags were introduced to distinguish the status of a point in different stages of the parallel fast marching method. In particular, $\texttt{BAND}$ and $\texttt{KNOWN}$ points in the overlapping zones were labelled as $\texttt{NEW}$ when their values were obtained from the quadratic solver and $\texttt{OLD}$ when they were collected to be sent to neighboring subdomains, respectively. This was beneficial for reducing the data size involved in communications as well as repetitive computations for the distributed-memory parallelization with overlapping domain decomposition. In this work, thanks to the shared-memory setting, such a distinguishment is found to be generally indifferent in terms of performance. Therefore, we only separate the $\texttt{KNOWN}$ category into two subsets:  a) $\texttt{KNOWN\_FIX}$, which contains the points initialized as boundary condition with fixed function values during the solution process; and b) $\texttt{KNOWN\_NEW}$, which contains the rest $\texttt{KNOWN}$ points defined in the original sequential fast marching method. This tag augmentation is still necessary, because a $\texttt{KNOWN\_NEW}$ point in one subdomain may require recomputation due to incoming characteristics from  neighboring subdomains \cite{yang2017highly}. Likewise, the boundary points are labeled as $\texttt{KNOWN\_FIX}$ instead of merely $\texttt{KNOWN}$ in the heap initialization step as shown in Algorithm \ref{alg:init_heap}. 

In this work, a modification of the heap initialization step given in \cite{yang2017highly} is that the boundary points labeled as $\texttt{KNOWN\_FIX}$ are also inserted into the heap now, showing as replacing $\texttt{Update\_Neighbors}$ with $\texttt{Insert\_Heap}$ in Algorithm \ref{alg:init_heap}. Correspondingly, it is necessary to verify that the top element in the heap is not in the $\texttt{KNOWN\_FIX}$ set before labeling it as $\texttt{KNOWN\_NEW}$ in Algorithm \ref{alg:march_new}. Therefore,  regardless of their status all points are put into the heap. Such a unified heap treatment slightly increases the computational cost due to filling the initial heap with some $\texttt{KNOWN\_FIX}$ elements. As a direct result, however, $\texttt{Update\_Neighbors}$ will be called only once and the code structure can be much simplified when block activation mechanisms are involved later. Moreover, with the small changes up to this point the whole algorithm still strictly follow the original sequential algorithm; that is also why we treat them separately in this part.

\begin{algorithm}
\algsetup{indent=1em}%
\caption{Solve the quadratic equation:\newline $\textsc{Solve\_Quadratic}(l,m,n{\color{green}{,ib}})$.}
\label{alg:quadratic}
\begin{algorithmic}[1]
\STATE $nd \leftarrow 0$
\STATE {Check the $x$ direction for upwind neighbor:}
  \STATE $d \leftarrow 0$
    \IF {$G ^{\color{green}{ib}} _{l-1,m,n}  \in \texttt{KNOWN}$ $\color{green}{\mathbf{and}\; \psi ^{ib} _{l,m,n}  > \psi ^{ib} _{l-1,m,n}  }$}
      \STATE $d \leftarrow -1$
    \ENDIF
    \IF {$G ^{\color{green}{ib}} _{l+1,m,n} \in \texttt{KNOWN}$ $\color{green}{\mathbf{and}\;  \psi ^{ib} _{l,m,n}  > \psi ^{ib} _{l+1,m,n}}$}
      \IF {$d = 0$ \OR $\psi  ^{\color{green}{ib}} _{l+1,m,n} < \psi  ^{\color{green}{ib}} _{l-1,m,n}  $}
        \STATE $d \leftarrow 1$
      \ENDIF
    \ENDIF
  \IF{$d \neq 0$}
    \STATE $nd \leftarrow nd + 1$
    \STATE $\psi_{nd} \leftarrow \psi  ^{\color{green}{ib}} _{l+d,m,n} $
  \ENDIF
\STATE {Check the $y$ direction for upwind neighbor:}
\STATE {$ \cdots \cdots $}
\STATE {Check the $z$ direction for upwind neighbor:}
\STATE {$ \cdots \cdots $}
\STATE {Reorder $ \psi_1 ,\cdots ,\phi_{nd}$ such that $ \psi_1 \le \cdots \le \phi_{nd}$}
\STATE $id \leftarrow   1$
\WHILE {$id \le nd$}
  \STATE $a \leftarrow  id$
  \STATE $b \leftarrow \sum_{i=1} ^{id} \psi_i $
  \STATE $c \leftarrow     \sum_{i=1} ^{id} \psi_i ^2  - \frac{ \Delta h ^2} {F_{l,m,n}^2 } $
  \IF {$(b^2-ac) \geq 0$}
    \STATE $\psi_\text{temp} \leftarrow \frac{b+\sqrt{b^2-ac}}{a}$
    \IF {$id < nd$ \AND $\psi_\text{temp}  > \psi_{id+1}$}
      \STATE $ id \leftarrow  id + 1 $
    \ELSE 
      \RETURN $\psi_\text{temp} $
    \ENDIF
  \ENDIF
\ENDWHILE
\end{algorithmic}
\end{algorithm}

\subsection{Block-based approach}\label{sec:nbfmm}

\begin{figure}[htbp!]
\begin{center}
 \includegraphics[angle=0,width=0.45\textwidth]{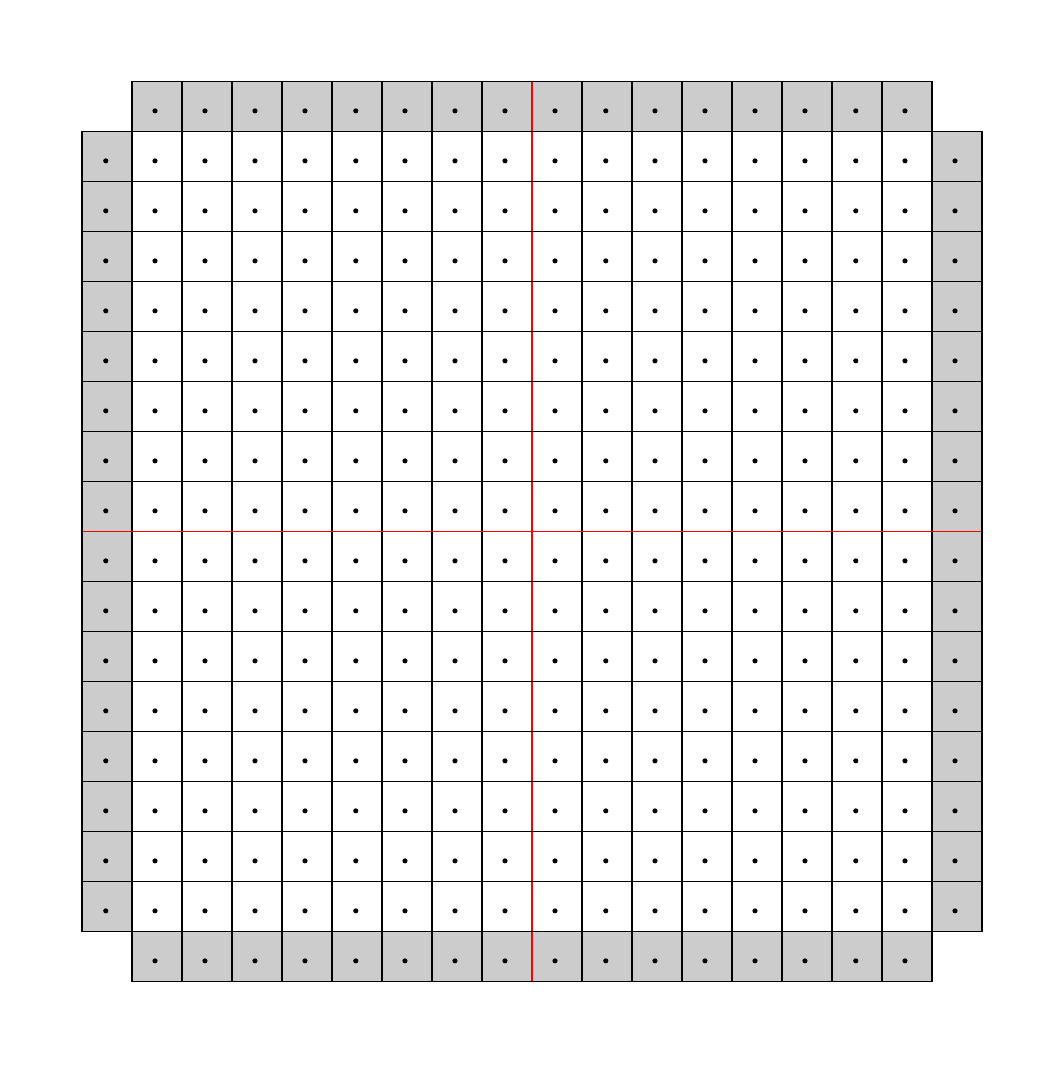}
 \includegraphics[angle=0,width=0.45\textwidth]{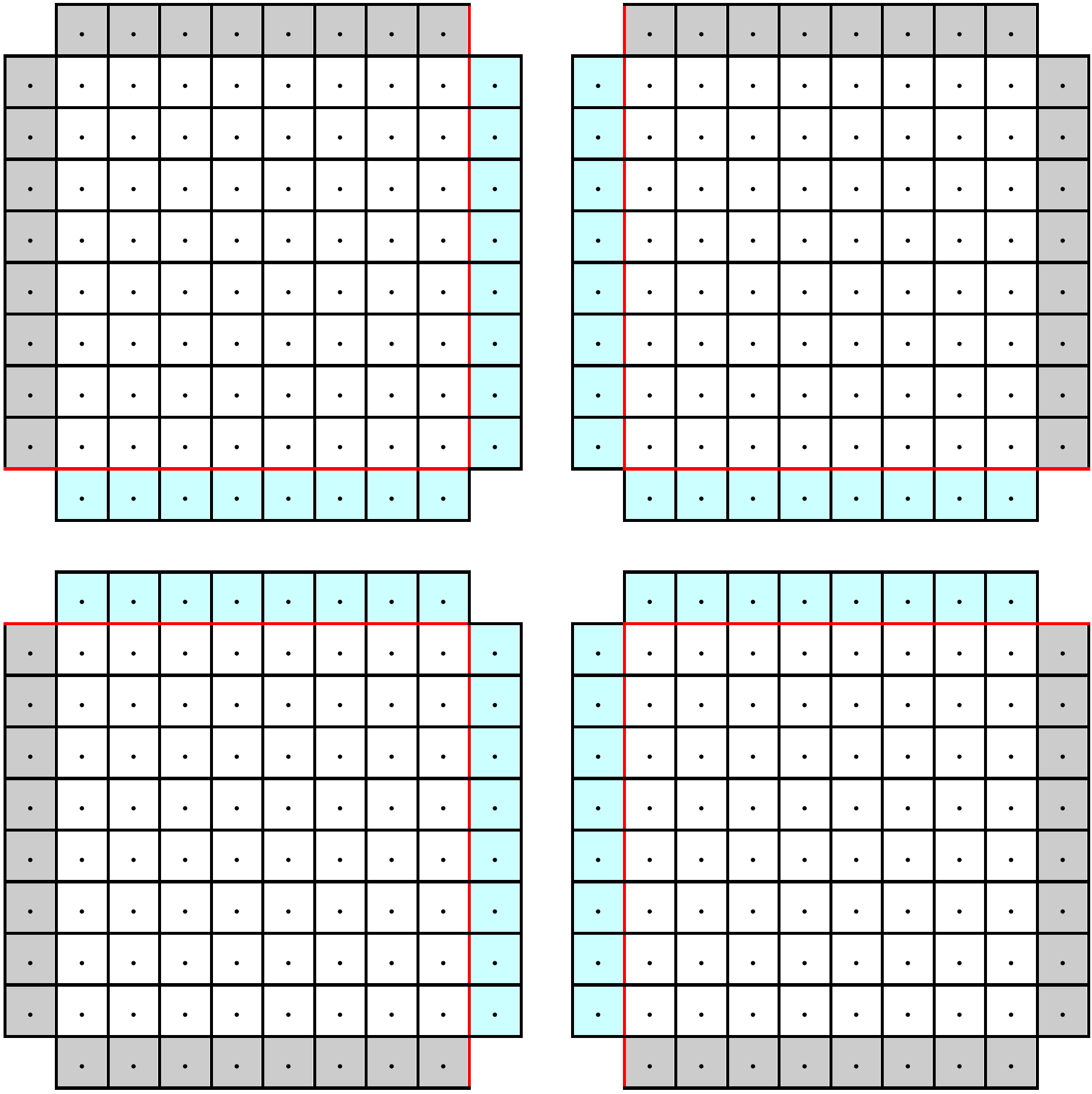}\\
\makebox[0.45\textwidth]{(a)}
\makebox[0.45\textwidth]{(b)} \\
\end{center}
 \caption{Computational domain: (a) single domain with ghost cell zones; (b) decomposed domain with ghost cell zones for data exchanges between non-overlapping subdomains.}
\label{fig:domain}
\end{figure}

Here we introduce a simple block-based approach for the fast marching method. For simplicity, each direction is evenly divided into $\texttt{mBlocks}$ pieces, which gives a total of $\texttt{nBlocks} = \texttt{mBlocks} ^3$ subdomains. As shown in Fig. \ref{fig:domain}(b) for a 2D case,  each subdomain are patched with ghost cell zones in both the lower and upper ends of each direction, just like the undecomposed domain in Fig. \ref{fig:domain}(a). For an arbitrary subdomain $ib$, it will have its own heap, which can be initialized and operated independently. Although $\psi$ and $G$ can be defined globally in a shared-memory setting, the algorithm on each subdomain follows the original sequential algorithm closer with locally defined variables. Moreover, a thread only handles a grid block instead of the whole domain at any time. With a limited amount of cache memory, the cache miss rate would be much lower for data structures of smaller size, hence the cache performance could be improved too. 
Therefore, as shown in the Algorithms discussed above, we add superscript $ib$ to $\varOmega$, $\varTheta$,  $\psi$, $G$, and the heap $\mathfrak{H}$ to emphasize that they are defined locally in each subdomain. 

Algorithm \ref{alg:pfmm} gives the main function for the block-based narrow band fast marching method. Apparently, all function
calls are embedded in the $\textbf{for}$ loop in order to be executed in each subdomain. As a simple way to run the sequential algorithm to obtain correct results, Algorithm \ref{alg:march_new} is placed in an infinite loop. The major addition in this loop not discussed above is the function that accesses the data of neighboring subdomains and integrates them into the heap if applicable. For convenience, let $\mathfrak{N}^{ib}$ represent the six neighboring blocks of block $ib$ (left, right, front, back, bottom, and top) and $\Pi ^{ib}$ represent the collection of points that are corresponding to the ghost cell zones of its neighboring blocks. Therefore, as shown in Algorithm \ref{alg:collect}, $\varTheta^{ib} \cap \Pi ^{nb}$ represents the portion of ghost cells in block $ib$ that are within the physical domain of block $nb$, If a point from $nb$ have a smaller function value, then its value and status can be copied to the corresponding point in block $ib$; this point will be added to $\mathfrak{H} ^{ib}$ or its position in $\mathfrak{H} ^{ib}$ will be adjusted for a current heap element.

It is imperative that the infinite loop has a proper exit condition. Here we introduce a logical variable $\mathfrak{A}$  to indicate whether a subdomain $ib$ is active for further marching work ($\mathfrak{A} ^{ib} =\texttt{TRUE}$) or not ($\mathfrak{A} ^{ib} = \texttt{FALSE}$). As shown in Algorithm \ref{alg:init_heap}, $\mathfrak{A} ^{ib}$ is initialized as $\texttt{FALSE}$ and set to be $\texttt{TRUE}$ if the subdomain has an nonempty initial heap. Then in Algorithm \ref{alg:march_new}, same as the exit conditions for the sequential algorithm, when the heap is empty or the top element in the heap is beyond $\texttt{width}_{\text{band}}$, this block should be set inactive. On the other hand, a block has to be set active if its heap has any elements less than the terminal bound after integrating the data from neighboring blocks through ghost cells, as shown at the end of Algorithm \ref{alg:collect}. 

It is also required to include the $\texttt{KNOWN\_NEW}$ neighbors of a $\texttt{KNOWN}$ point just removed from the heap in the check for possible updates, i.e., only $\texttt{KNOWN\_FIX}$ neighbors can be excluded from the check as shown in Algorithm \ref{alg:update}. In addition, the quadratic equation is only solved at a neighbor with a larger value. Similar checks are imposed on looking for upwind points in Algorithm \ref{alg:quadratic}. These comparisons of function values added here for bypassing unnecessary computations would be redundant in the sequential algorithm, because points added to the $\texttt{KNOWN}$ set follow a strict order of increasing function values if only a single domain is involved. As discussed above and in \cite{yang2017highly} with more details, the sequential algorithm has not been altered by the introduction of the augmented status tags and these checks. However, they make the block-based algorithm a very natural extension of the sequential one.

\subsection{Restarted narrow band approach} \label{sec:restarted}

With the above block-based approach, it is clear that the sequential fast marching algorithm will run in each block until exit, then data will be exchanged among neighbors via ghost cells; and both the marching and exchanging steps repeat until the global exit condition is met. This is straightforward in concept and implementation, but the performance is unsatisfying by any means. In \cite{yang2017highly}, a novel restarted narrow band approach was proposed to address the performance issue by replacing the terminal narrow band bound $\texttt{width}_{\text{band}}$ with an evolving one, i.e., $\texttt{bound}_{\text{band}}$ in Algorithm \ref{alg:march_new}. It should be noted here that the condition to label the block as inactive is still determined by $\texttt{width}_{\text{band}}$. On the other hand, as shown in Algorithm \ref{alg:pfmm}, a free parameter $\texttt{stride}$ is introduced to advance the front by the extent of $\texttt{stride}$, or $\delta s$, in each run-through.  Apparently, with $\texttt{stride} = \texttt{width}_{\text{band}}$ the previous block-based approach is fully restored.

\begin{algorithm}
\algsetup{indent=1em}
\caption{Front propagation within the narrow band:\newline $\textsc{March\_Narrow\_Band}{\color{green}{(ib)}}$.}
\label{alg:march_new}
\begin{algorithmic}[1]
    {\color{magenta}{\STATE {$\Upsilon ^{ib} \leftarrow \varnothing$}}}
    \LOOP 
      \IF {$\texttt{size} (\mathfrak{H}^{\color{green}{ib}} ) = 0$}
       {\color{green}{ \STATE $\mathfrak{A} ^{ib} \leftarrow  \texttt{FALSE}$}}
        \STATE \textbf{exit loop}
      \ENDIF
      \STATE $(i,j,k) \leftarrow  \textsc{Locate\_Min}(\mathfrak{H}^{\color{green}{ib}} )$  
      \IF{$\psi ^{\color{green}{ib}} _{i,j,k}  > \blueout{\texttt{width}_\text{band}}\; \color{blue}{\texttt{bound}_\text{band}} $}
      {\color{blue}{\IF {$\psi ^{ib} _{i,j,k}   > \texttt{width}_\text{band}$}
       {\color{green} {\STATE $\mathfrak{A} ^{ib} \leftarrow  \texttt{FALSE}$}}
      \ENDIF}}
        \STATE \textbf{exit loop}
      \ENDIF
     {\color{red}{\IF{$G ^{\color{green}{ib}} _{i,j,k} \notin \texttt{KNOWN\_FIX}$}
        {\color{black}{\STATE $G ^{\color{green}{ib}} _{i,j,k}  \leftarrow  \texttt{KNOWN{\color{red}{\_NEW}}}$}}
      \ENDIF}}
     \STATE $\textsc{Remove\_Min}(\mathfrak{H}^{\color{green}{ib}}) $  
      \STATE $\textsc{Update\_Neighbors}(i,j,k{\color{green}{,ib}})$
    \ENDLOOP
  {\color{magenta}{\FORALL{block $nb \in \mathfrak{N} ^{ib} $ such that $\varTheta ^{nb} \cap \Upsilon ^{ib}  \neq \varnothing$}
           \STATE $\mathfrak{C} ^{nb} (ib)   \leftarrow  \texttt{TRUE}$
  \ENDFOR}}
\end{algorithmic}
\end{algorithm}

Unlike \cite{yang2017highly}, in which $\texttt{bound}_{\text{band}}$ was built on the smallest one among all top elements of the local heaps, here we choose a much simpler way to increase $\texttt{bound}_{\text{band}}$ evenly with $\texttt{stride}$ in each run-through. This simplification renders the global reduction operation of top heap elements unnecessary. More importantly, the second marching step in \cite{yang2017highly} is not required here, which makes the restarted narrow band approach become not only more compact and more efficient, but also more consistent with the sequential algorithm. 

\subsection{Mechanism for activating  the marching step} \label{sec:marching}

Due to the narrow band nature of the fast marching method, the computational load is unevenly distributed in space during different stages of the front marching process. With the current block-based, restarted narrow band approach, it is conceivable that some blocks will be free of  load, either because the block has been fully updated (i.e., the local heap is exhausted,  or the terminal bound  $\texttt{width}_\text{band}$ is surpassed by the top element of the local heap) or because the moving front has not reached this block yet. In this case,  as shown in Algorithm \ref{alg:march_new}, the function call of the marching step will  quit immediately without doing any real work. Therefore, a simple mechanism is introduced in Algorithm \ref{alg:pfmm} such that the marching step for a block is activated only if $\mathfrak{A}  ^{ib} = \texttt{true}$. Clearly this precondition won't change the results, but the overheads of repeated function calls can be avoided. It should also be noted that the only place beyond the heap initialization that $\mathfrak{A}  ^{ib}$ could be assigned $\texttt{true}$ is in Algorithm \ref{alg:collect} as discussed earlier. 

\begin{algorithm}
\algsetup{indent=1em}
\caption{Integrate data in the ghost cell region:\newline $\color{green}{\textsc{Integrate\_Ghost\_Cell\_Data}(ib)}$.}
\label{alg:collect}
\begin{algorithmic}[1]
 {\color{green}{ \FORALL{block $nb$ such that $nb \in \mathfrak{N}^{ib} $  $\color{magenta}{\mathbf{and}\; \mathfrak{C}^{ib} (nb)  =  \texttt{TRUE}}$} 
  \FORALL{$(i,j,k)$ such that $(i,j,k) \in  \varTheta^{ib} \cap \magentaout{\Pi ^{nb}}\; \color{magenta}{\Upsilon^{nb}}$}
    \IF{$\psi_{i,j,k} ^{ib} > \psi_{i,j,k} ^{nb}$}
        \STATE {$\psi_{i,j,k} ^{ib} \leftarrow  \psi_{i,j,k} ^{nb} $}
        \STATE {$G_{i,j,k} ^{ib} \leftarrow  G_{i,j,k} ^{nb}$}
        \IF{$(l,m,n) \not\in \mathfrak{H} ^{ib} $} 
            \STATE {$\textsc{Insert\_Heap}(l,m,n,\mathfrak{H} ^{ib})$ }
        \ELSE 
            \STATE {$\textsc{Up\_Heap}(l,m,n,\mathfrak{H} ^{ib})$ }
        \ENDIF
      \ENDIF
    \ENDFOR
\ENDFOR

        {\color{magenta}{\STATE {$\mathfrak{C} ^{ib} \leftarrow  \texttt{FALSE}$}}}

      \IF {$\texttt{size} (\mathfrak{H}^{ib} ) \neq 0 $}
      \STATE $(i,j,k) \leftarrow  \textsc{Locate\_Min}(\mathfrak{H}^{ib})$  
      \IF {$\psi ^{ib} _{i,j,k}  \leq \texttt{width}_\text{band}$}
        \STATE $\mathfrak{A} ^{ib} \leftarrow  \texttt{TRUE}$
        \ENDIF
      \ENDIF}}
\end{algorithmic}
\end{algorithm}

\subsection{Mechanism for activating the exchanging step} \label{sec:exchanging}

Designing a mechanism for activating the exchange step is slightly more involved because each block has six neighbors. Here we introduce a six-element logical variable $\mathfrak{C}$ for each block to indicate possible incoming characteristics from neighboring blocks. As shown in Algorithm \ref{alg:pfmm}, $\mathfrak{C} ^{ib}$ is initialized as $\texttt{FALSE}$ at the beginning. The exchanging step is required for a block if any of the six elements becomes $\texttt{TRUE}$ in the marching step, i.e., Algorithm \ref{alg:march_new}. As mentioned earlier, $\Pi ^{ib} $ represents the set of points in block $ib$ that correspond to the ghost cells in the neighbors of block $ib$. Initialized to be empty, $\Upsilon ^{ib}$ is the set of points within $\Pi ^{ib} $ that are updated through the quadratic solver and collected in Algorithm \ref{alg:update}. If any points in $\Upsilon ^{ib}$ and the ghost cells of a neighboring block $nb$ coincide, the element of $\mathfrak{C} ^{nb}$ corresponding to block $ib$ should be set to $\texttt{TRUE}$. Then in Algorithm \ref{alg:collect}, a neighboring block $nb$ will be used to possibly update the ghost cells of block $ib$ only if the corresponding element of $\mathfrak{C} ^{ib}$ has been set to $\texttt{TRUE}$ in the marching step. Similarly, $\Pi ^{ib} $ can be replaced by $\Upsilon ^{ib}$ to further limit the comparison operations.  If should be noted that $\mathfrak{C} ^{ib}$ is reset to $\texttt{FALSE}$ after all possible ghost cell updates for block $ib$ have been checked.

\subsection{Shared-memory parallelization}\label{sec:shared}

It is evident that Algorithm \ref{alg:pfmm} mainly consists of three $\textbf{for}$ loops. Therefore, a shared-memory parallelization is simply to run these three loops with multiple threads. This can be easily achieved by using the $\textbf{parallel do}$ directive in Fortran or the $\textbf{parallel for}$ directive in C/C++ to convert them into parallel regions.

It might be surprising that there is no race condition in the current shared-memory parallel algorithm. As mentioned earlier, $\psi$, $G$, and the heap $\mathfrak{H}$ are defined locally in each block. In Algorithm \ref{alg:collect}, the thread working on block $ib$ will access the points within its neighboring block $nb$ to possibly update the ghost cells of block $ib$. Because of the non-overlapping domain decomposition adopted in this work, the operations on the ghost cells in the exchange step are totally isolated from those on physical cells within the block in the marching step. On the other hand, the thread working on block $ib$ will also access $\mathfrak{C} ^{nb}$ of the neighboring block $nb$ in Algorithm \ref{alg:march_new}. Again, this won't make it a race condition as only the element corresponding to $ib$ in $\mathfrak{C} ^{nb}$ could be changed.

\begin{algorithm}
\algsetup{indent=1em}
\caption{Block-based, restarted narrow-band fast marching method:\newline $\textsc{{\color{green}{Block\_Based\_}}{\color{blue}{Restarted\_}}Narrow\_Band\_Fast\_Marching}$.}
\label{alg:pfmm}
\begin{algorithmic}[1]
{\color{blue}{ \STATE $\texttt{bound}_\text{band} \leftarrow 0$ }}
   {\color{magenta}{\STATE $\mathfrak{C}  \leftarrow  \texttt{FALSE}$}}
{\color{green}{   
      \FORALL{block $ib$ such that $ 1 \leq ib \leq nblocks$}
           {\color{black}{\STATE $\textsc{Initialize\_Heap}{\color{green}{(ib)}}$}}
      \ENDFOR

    \LOOP 

{\color{blue}{        
        \STATE $\texttt{bound}_\text{band}  \leftarrow  \min \left( \texttt{bound}_\text{band}  + \texttt{stride}, \texttt{width}_\text{band} \right)$}}
      \FORALL{block $ib$ such that $ 1 \leq ib \leq nblocks$ {\color{cyan}{$\mathbf{and} \; \mathfrak{A}  ^{ib} = \texttt{TRUE}$}}}
{\color{black}{        
          \STATE $\textsc{March\_Narrow\_Band}{\color{green}{(ib)}}$ }}
      \ENDFOR

      \FORALL{block $ib$ such that $ 1 \leq ib \leq nblocks$ {\color{magenta}{$\mathbf{and} \; \texttt{any}(\mathfrak{C} ^{ib} = \texttt{TRUE})$}}}
          \STATE $\textsc{Integrate\_Ghost\_Cell\_Data}(ib)$ 
      \ENDFOR

      \IF {$\texttt{all}(\mathfrak{A} = \texttt{FALSE})$}%
        \STATE \textbf{exit loop}
      \ENDIF

    \ENDLOOP}}
\end{algorithmic}
\end{algorithm}

\section{Results}\label{sec:benchmark}

\subsection{Test cases}

Four point source problems tested in \cite{yang2017highly} were performed. For simplicity, a unit cube $[-0.5,0.5] \times [-0.5,0.5] \times [-0.5,0.5]$ was used as the physical domain and the point source was set at the domain center. Three cases were derived from \cite{ChaconV2015} with speed functions defined by 
\begin{eqnarray}
F(x,y,z) &=& 1, \label{eq:f1}\\  
F(x,y,z) &=& 1 + 0.50 \sin(20 \pi x)  \sin(20\pi y)  \sin(20\pi z), \label{eq:f2}\\  
F(x,y,z) &=& 1 - 0.99 \sin(2  \pi x)  \sin(2 \pi y)  \sin(2 \pi z), \label{eq:f3}
\end{eqnarray} 
respectively. For the last case,  in the domain of $F = 1$, there are four concentric spherical obstacles of $F = 0$ defined by 
\begin{equation}
\begin{array}{ccc}
(0.15 < R < 0.15+w) &\setminus & ((r<0.05) \cap (z < 0));\\
(0.25 < R < 0.25+w) &\setminus & ((r<0.10) \cap (z > 0));\\
(0.35 < R < 0.35+w) &\setminus & ((r<0.10) \cap (z < 0));\\
(0.45 < R < 0.45+w) &\setminus & ((r<0.10) \cap (z > 0)),
\end{array}
\label{eq:barriers}
\end{equation}
where $\setminus$ represents the set difference operation, $w = \tfrac{1}{24}$, $R = \sqrt{ x^2+y^2+z^2 }$, and $r= \sqrt{ x^2+y^2 }$. This case was derived from \cite{DetrixheGM2013}.

Four grids of uniform spacing $\Delta h$ in all three directions were used and the number of points in each direction (excluding the ghost points) was $ \texttt{dimGrid} = 128$, $256$, $512$, and $1024$, respectively. Eight block sizes were tested, in which the number of grid points in each direction of the block was $ \texttt{dimBlock} = 8$, $16$, $32$, $64$, $128$, $256$, $512$, and $1024$, respectively. Therefore, the partition in each direction is the same for every configuration and is simply labelled as $\texttt{mBlocks}$ here. The total number of blocks for the whole domain is $\texttt{nBlocks} = \texttt{mBlocks} ^3$. Unlike the stationary domain decomposition for distributed-memory parallelization in \cite{yang2017highly}, here the number of blocks is not restricted by the number of threads, $\texttt{nThreads}$, used for the computation as long as $\texttt{nThreads} \leq \texttt{nBlocks}$. Hence, the block size $\texttt{dimBlock}$ can be considered as a free parameter with a range of $[1,\texttt{dimGrid}]$ in this study. On the other hand, as a restarted narrow band approach, $\texttt{stride}$ introduced in \cite{yang2017highly} is still a free parameter in this work. And the selection of the stride size $\delta s$ also follows the test spectrum in \cite{yang2017highly}, i.e., $\delta s = 0.5 \Delta h$, $1.5 \Delta h$, $2.0 \Delta h$, $2.5 \Delta h$, $3.5 \Delta h$, and $\infty$. 

All computations were performed on two desktop computers. The first computer is equipped with a 16-core/32-thread AMD Ryzen Threadripper 1950X  3.4 GHz processor and 64 GBytes of DDR4 memory. The second computer is equipped with a 64-core/256-thread Intel Xeon Phi 7210 1.3 GHz processor and 96 GBytes of DDR4 memory. In this study, the algorithm was implemented in Fortran 2003. The code was compiled in double precision using the Intel Fortran Compiler XE version 18.0.0 with the $\texttt{-O3}$ optimization option. Compiling options $\texttt{-qopenmp}$ and $\texttt{-qopenmp-stubs}$ ware used to generate the multi-threaded parallel program and the sequential program, respectively. The sequential algorithm described in Sec.  \ref{sec:sfmm} with the modifications in Sec. \ref{sec:uniheap} was considered as the baseline algorithm for comparison. 

In summary, four example problems were carried out on four grids. The grids were decomposed using eight different block sizes as applicable. Three solvers were applied to these test cases: (a) baseline (B) solver, which can treat the whole domain only as a single block with a single process; (b) block-based sequential (S) solver, which can handle a decomposed domain of different block sizes using a single process; and (c) block-based parallel (P) solver, which can handle a decomposed domain of different block sizes using multiple threads via OpenMP directives. The latter two are also labeled as M because both of them can handle multiple blocks. 

\subsection{Accuracy}

\begin{figure}[htbp!]
\begin{center}
 \includegraphics[angle=0,width=0.4\textwidth]{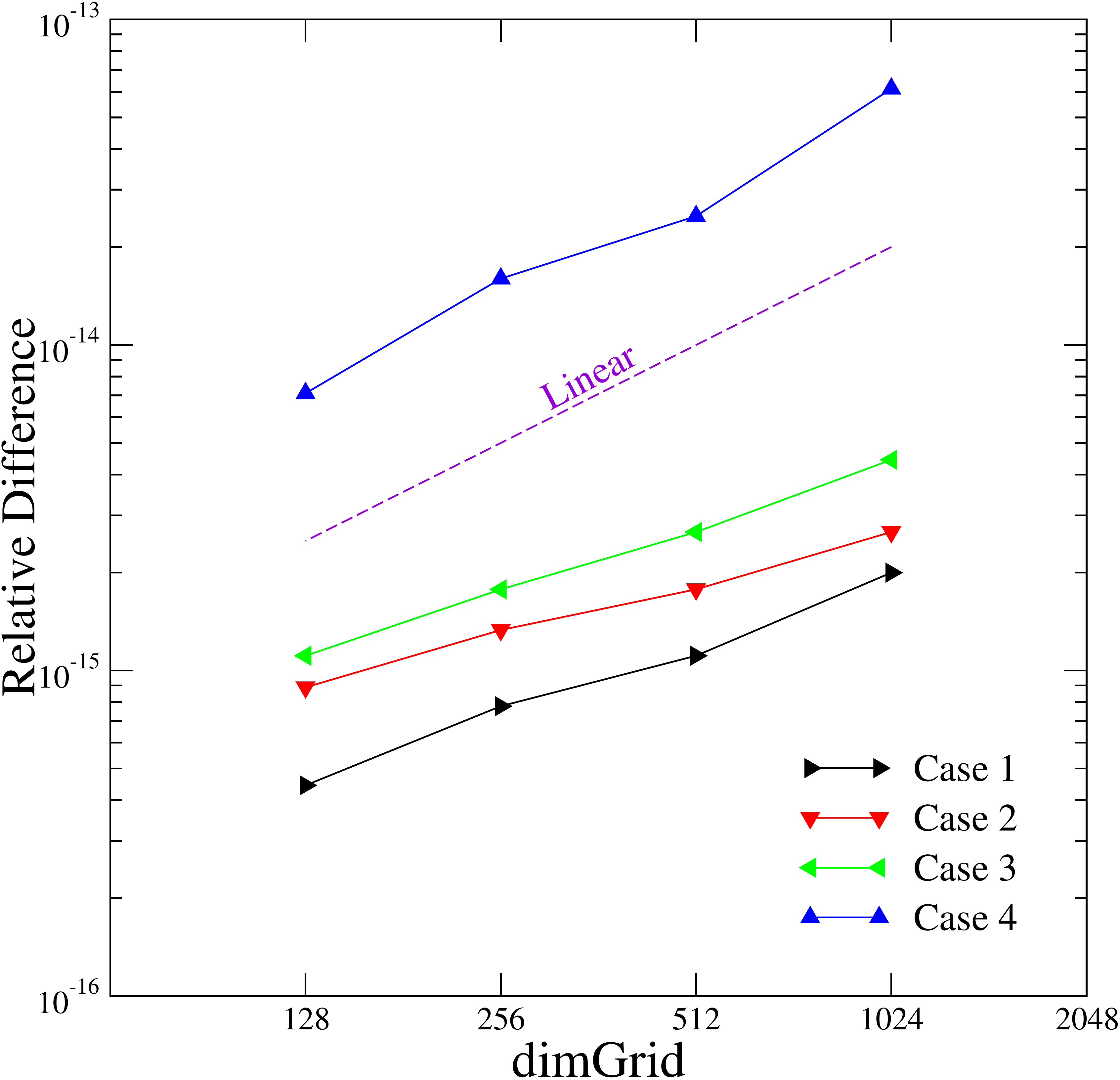}
\end{center}
 \caption{The maximum relative differences between the multi-block and single-block solutions.}
\label{fig:differences}
\end{figure}

A comparative study was first performed to verify that the block-based algorithms give the correct solutions. For each problem, a benchmark solution ($\psi ^B$) can be obtained on each grid using the baseline sequential solver. The maximum relative difference between this solution and the solutions from the other two solvers ($ \psi  ^M$) is defined as $ \psi^{rd} = \max_{i,j,k} |\psi _{i,j,k} ^B - \psi _{i,j,k} ^M | / \psi _{i,j,k} ^B  $. Note that the solutions using a single block from all three solvers are exactly the same. Therefore, the maximum value of $ \psi^{rd}$ from all multi-block solutions on the same grid is obtained. As shown in Fig. \ref{fig:differences}, it is evident that for all cases the differences are within the order of machine accuracy for double precision floating point calculations. Also, as the grid refines, the relative error grows following a linear or sub-linear rate. Both phenomena are consistent with what observed in \cite{yang2017highly}.

\subsection{Overheads}

\begin{figure}[htbp!]
\begin{center}
 \includegraphics[angle=0,width=0.95\textwidth]{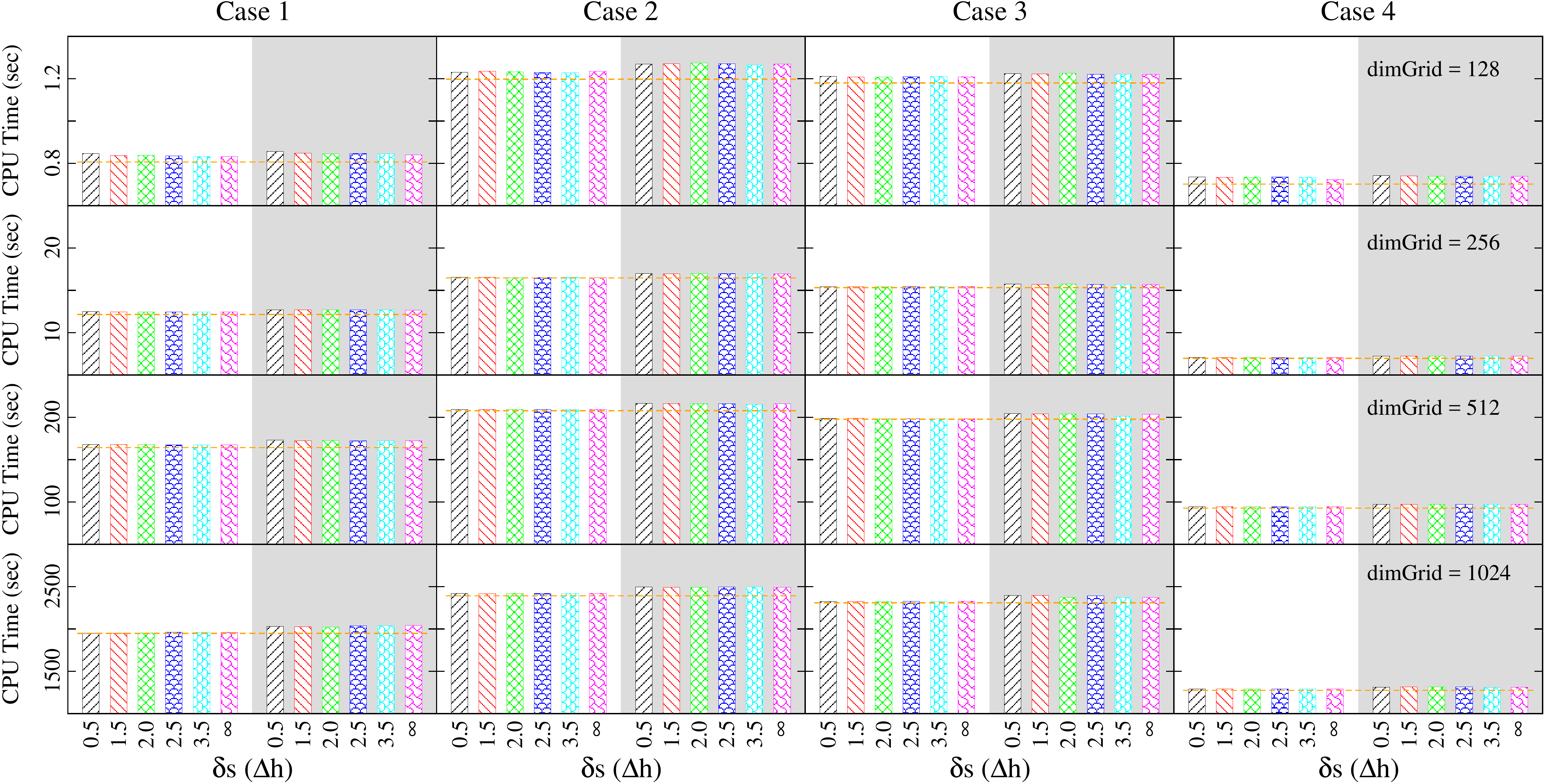}
 \end{center}
 \caption{Single-block runs on AMD Ryzen Threadripper: the CPU time as a function of the stride size $\delta s$ and the grid size $\texttt{dimGrid}$. The horizontal orange lines represent $T_B$ from the baseline sequential solver. The left part (blank background) and right part (grey-shaded background) show $T_S$ and $T_P$ from the multi-block sequential and parallel solvers, respectively. }
\label{fig:one-block-cputime}
\end{figure}
\begin{figure}[htbp!]
\begin{center}
 \includegraphics[angle=0,width=0.95\textwidth]{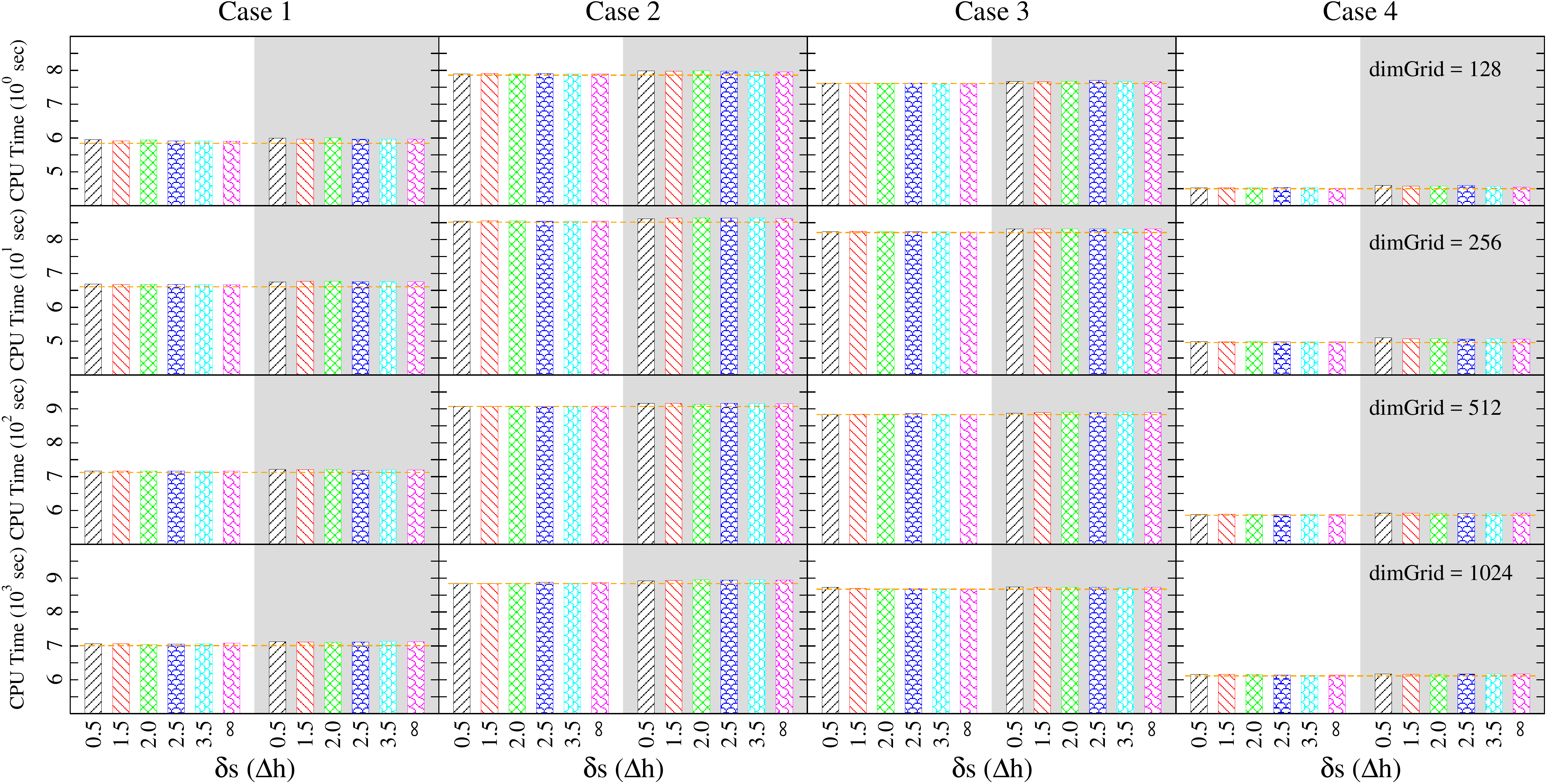}
 \end{center}
 \caption{Single-block runs on Intel Xeon Phi: the CPU time as a function of the stride size $\delta s$ and the grid size $\texttt{dimGrid}$. The horizontal orange lines represent $T_B$ from the baseline sequential solver. The left part (blank background) and right part (grey-shaded background) show $T_S$ and $T_P$ from the multi-block sequential and parallel solvers, respectively.}
\label{fig:one-block-cputime-phi}
\end{figure}

In addition to the floating point differences in the solutions, another aspect of the multi-block solvers can be checked is the algorithm overheads extra to the baseline solver. Unlike the distributed-memory parallel algorithm in \cite{yang2017highly}, in which all functions extra to the marching step were called and executed, here in a single-block setting Algorithm \ref{alg:collect} won't even be called in the two multi-block solvers. Therefore, the overheads are mainly from setting up loops in Algorithm \ref{alg:pfmm} and $\texttt{if}$ checks in Algorithms \ref{alg:march_new} and \ref{alg:update}. Of course, using OpenMP threading also involves additional overheads such as OpenMP library startup, thread startup, loop scheduling, etc. Especially, our current OpenMP parallelization was implemented by tackling three separate loop regions (including one initialization loop) using the OpenMP parallel directive, so the thread startup overhead is always there in each restart. However, as discussed earlier,  there is no race condition at all in our current parallel algorithm, hence it does not involve lock management, which usually is the major performance penalty in thread parallelization.

Figures \ref{fig:one-block-cputime} and \ref{fig:one-block-cputime-phi} show the CPU times from the single-block runs of three solvers on the AMD Ryzen Threadripper and Intel Xeon Phi computers, respectively. In general, the former is a few times faster than the latter (e.g., more than 7 times for small $\texttt{dimGrid}$, and less than 4 times for large $\texttt{dimGrid}$), but the overall trends of the results from the two computers are very similar. On the former, the CPU time $T_S$ can show up to 5\% overhead with the coarsest grid and less than 1\% overhead with the finest grid, whereas on the latter, $T_S$ generally gives an overhead less than 1\%. With the parallel solver, $T_P$ shows roughly around 1\% extra overhead above $T_S$ due to the extra OpenMP threading overheads discussed above. On the other hand, case 2 is the most expensive one because of its fast varying speed function, case 3 is slightly less expensive than case 2 as its speed function is smoother than that of case 2, and case 4 requires the least CPU time since the quadratic equation was not solved on the grid points within the barriers. 

\subsection{Restarts}

\begin{figure}[htbp!]
\begin{center}
 \includegraphics[angle=0,width=0.95\textwidth]{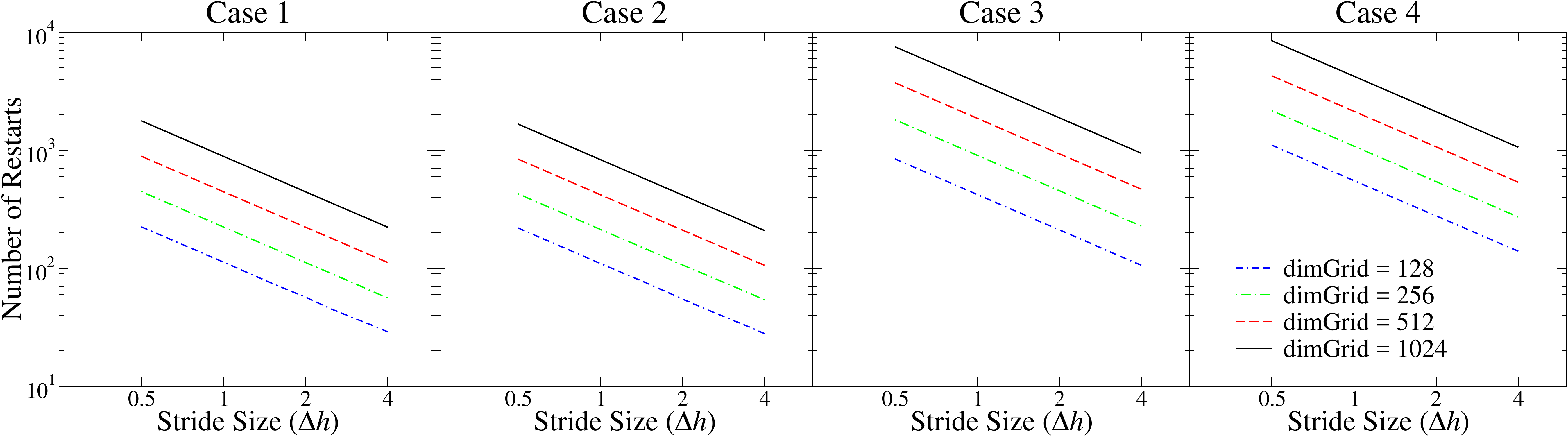}
\end{center}
 \caption{The number of restarts $nr$ as a function of the stride size $\delta s$ (from $\delta s = 0.5\Delta h$ to $\delta s = 3.5\Delta h$) and the grid size $\texttt{dimGrid}$.}
\label{fig:stride}
\end{figure}

As shown in Figure \ref{fig:stride}, the number of restarts $nr$ is clearly a linear function of the stride size $\delta s$ and the grid size $\texttt{dimGrid}$ for all test cases. A major difference between the results from the current block-based algorithm and those from the parallel algorithm in \cite{yang2017highly} is that the number of restarts shows no dependence on the domain decomposition configurations here. That is, for a given stride size on a given grid, no matter what is the number of partitions $\texttt{mBlocks}$, both multi-block solvers gave exactly the same number of restarts and there is no need to specify $\texttt{mBlocks}$ in the figure. As an explanation, the narrow band bound for each restart in the current algorithm is preset and evenly distributed. Therefore, unlike the original restart narrow band approach, it does not rely on the intermediate solution at all.

\begin{figure}[htbp!]
\begin{center}
 \includegraphics[angle=0,width=0.95\textwidth]{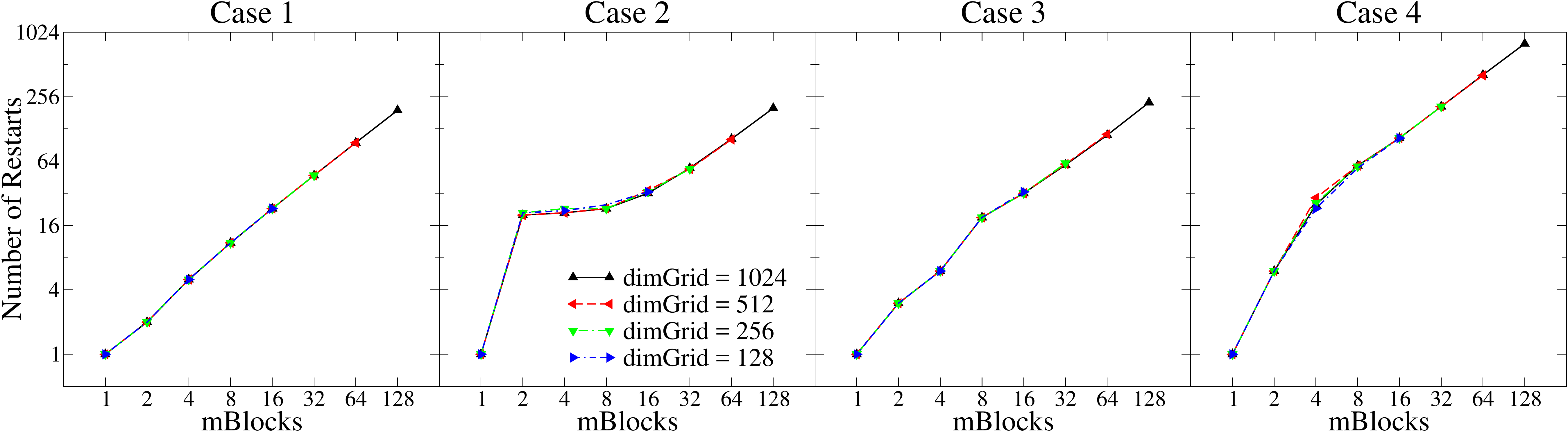}
\end{center}
 \caption{The number of restarts $nr$ as a function of  the number of partitions $\texttt{mBlocks}$ and the grid size $\texttt{dimGrid}$ for the stride size $\delta s = \infty$.}
\label{fig:stride2}
\end{figure}

 For the extreme case $\delta s = \infty$, Figure \ref{fig:stride2} shows the number of restarts $nr$ as a function of the grid size $\texttt{dimGrid}$ and the number of partitions $\texttt{mBlocks}$. It is evident that there is a very weak dependence of $nr$ on the grid size as the data from coarser grids almost collapse on those from the finer grids. In addition, the number of restarts asymptotically approaches a linear relationship with $\texttt{mBlocks}$ for all four problems. Although the choice of $\delta s = \infty$ is generally not recommended in practical applications of the current algorithm, the results here do verify that, unlike iterative methods, our new algorithm retains the non-iterative property of the original sequential fast marching method and it does not reply on any user-specified convergence criteria. 

\subsection{Sequential performance}

\begin{figure}[htbp!]
\begin{center}
  \includegraphics[angle=0,width=0.95\textwidth]{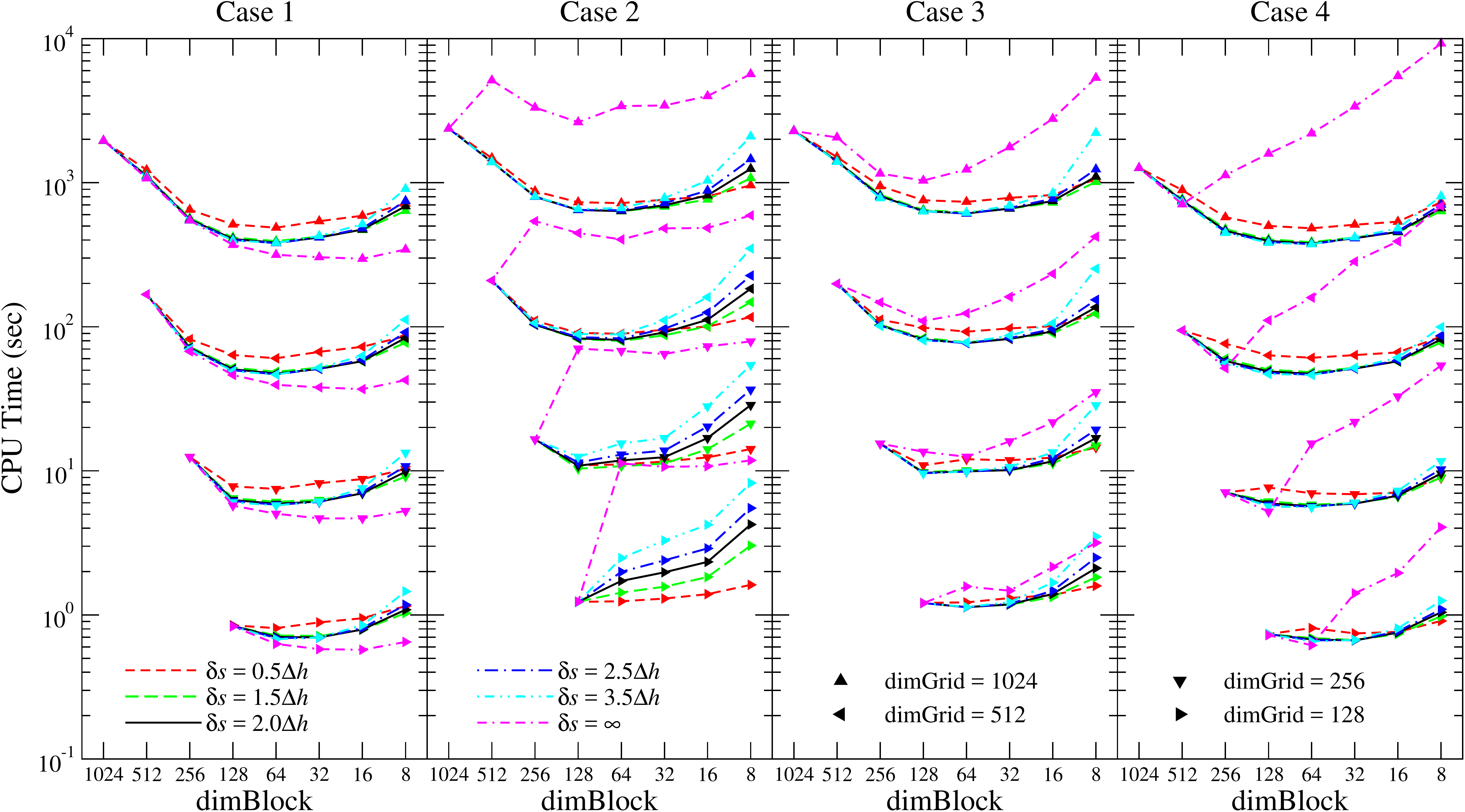}
\end{center}
 \caption{Multi-block sequential runs on AMD Ryzen Threadripper: the CPU time  $T_S$ as a function of the block size $\texttt{dimBlock}$, the stride size $\delta s$, and the grid size $\texttt{dimGrid}$. }
\label{fig:cputime_1t}
\end{figure}

\begin{figure}[htbp!]
\begin{center}
  \includegraphics[angle=0,width=0.95\textwidth]{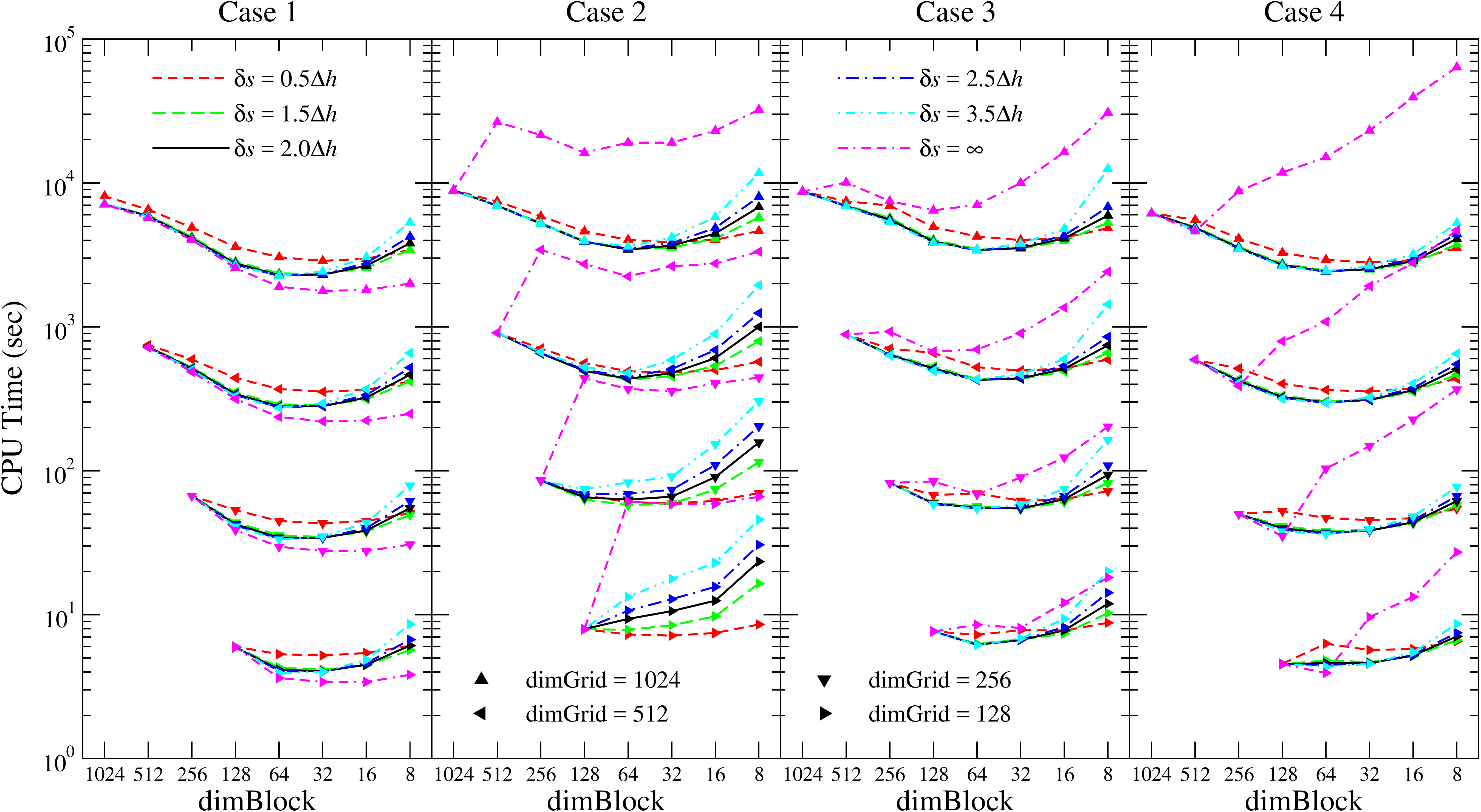}
\end{center}
 \caption{Multi-block sequential runs on  Intel Xeon Phi: the CPU time  $T_S$ as a function of the block size $\texttt{dimBlock}$, the stride size $\delta s$, and the grid size $\texttt{dimGrid}$. }
\label{fig:cputime_1t_phi}
\end{figure}

Figures \ref{fig:cputime_1t} and \ref{fig:cputime_1t_phi} show the CPU time $T_S$ as a function of the block size $\texttt{dimBlock}$, the stride size $\delta s$, and the grid size $\texttt{dimGrid}$ for the multi-block sequential solver on the AMD Ryzen Threadripper and the Intel Xeon Phi computers, respectively. A striking finding from the results is the much improved performance on the decomposed domain, especially, around $\texttt{dimBlock} = 64$ on finer grids. Only Case 2 on the coarsest grid, $\texttt{dimGrid} = 128$, didn't benefit from the domain decomposition. The results from both computers follow a very similar trend. In general, the improvements become more pronounced and the optimal $\texttt{dimBlock}$ converges to $64$ for all four cases as the grid refines. Interestingly, the choice of $\delta s = \infty$ gave the best performance for Case 1, which has a simple speed function of $F = 1$, but failed to compete with other values of $\delta s$ for other cases, except the $\texttt{mBlocks}=2$ configuration in Case 4 due to the special arrangement of the obstacles in the domain. This is a clear indication of the great merit of our restarted narrow band approach in general situations. On the other hand, the choice of $\delta s = 0.5 \Delta h$ performed better than others on grids $\texttt{dimGrid} = 128$  and $256$ for Case 2. Apparently more frequent data exchange among neighboring blocks works better for the situation of a fast changing speed function on a coarse grid. The results are not very sensitive to the difference in $\delta s$ around $\delta s = 2 \Delta h$. But similarly, an increased frequency of data exchange (i.e., a decreased stride size from $\delta s = 3.5 \Delta h$ to $1.5 \Delta h$) gives a better performance for a smaller $\texttt{dimBlock}$. 

\begin{figure}[htbp!]
\begin{center}
  \includegraphics[angle=0,width=0.95\textwidth]{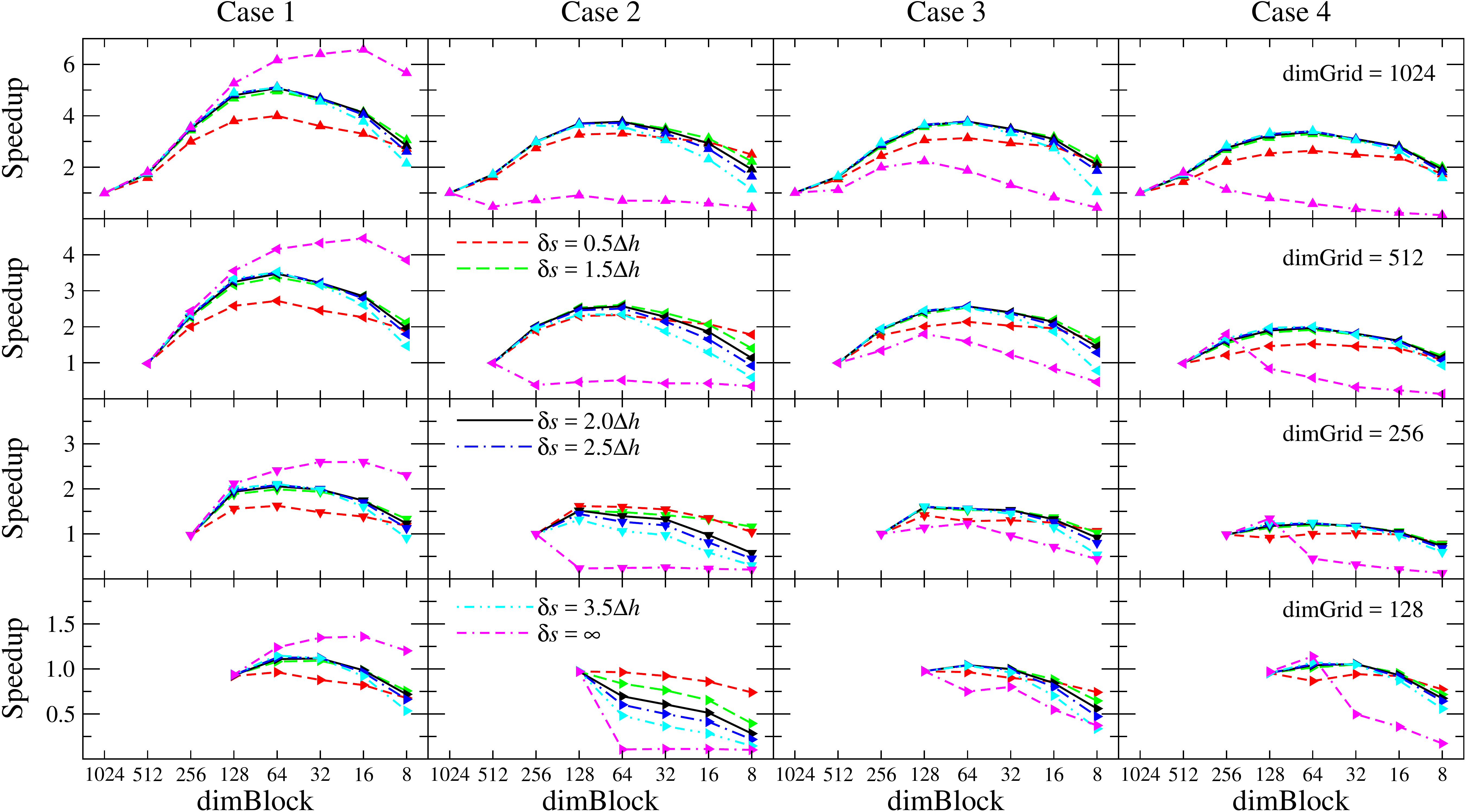}
\end{center}
 \caption{Multi-block sequential runs on AMD Ryzen Threadripper: the sequential speedup as functions of the block size $\texttt{dimBlock}$, the stride size $\delta s$, and the grid size $\texttt{dimGrid}$.}
\label{fig:speedup_1t}
\end{figure}

\begin{figure}[htbp!]
\begin{center}
  \includegraphics[angle=0,width=0.95\textwidth]{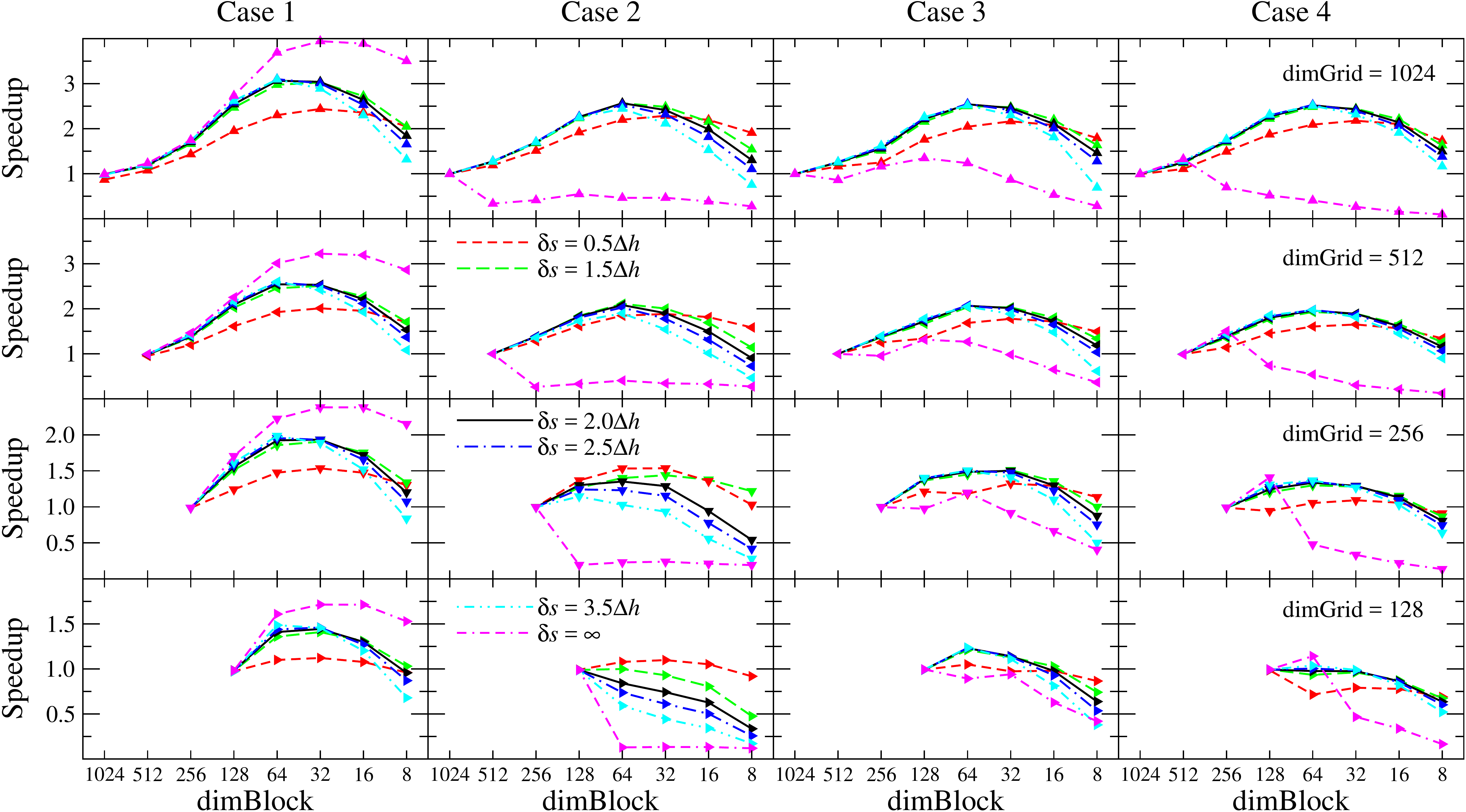}
\end{center}
 \caption{Multi-block sequential runs on  Intel Xeon Phi: the sequential speedup as functions of the block size $\texttt{dimBlock}$, the stride size $\delta s$, and the grid size $\texttt{dimGrid}$. }
\label{fig:speedup_1t_phi}
\end{figure}

The corresponding speedups of the multi-block sequential runs calculated using $T_B/T_S$ are shown in Figures \ref{fig:speedup_1t} and \ref{fig:speedup_1t_phi}.  The extreme option of $\delta s = \infty$ gave a very high speedup for Case 1, but performed poorly in other cases. As discussed above, it should not be considered as a general choice for our restarted narrow band approach in practical applications, hence it won't be included in the discussions hereafter except otherwise mentioned. On the AMD Ryzen Threadripper computer, the computation of Case 1 was only slightly accelerated on the grid $\texttt{dimGrid} = 128$, but it gave a speedup more than 2, 3, and 5 on the grid $\texttt{dimGrid} = 256$, $512$, and $1024$, respectively. The other three cases have more complex speed functions, and the highest speedups are relatively lower, e.g., around 3.5 on the finest grid. On the Intel Xeon Phi computer, the differences in performance among these cases are not so significant, although the speedup is above 3 for Case 1 and around 2.5 for other three cases on the finest grid. In general, the performance of the multi-block sequential solver on two different computers is very consistent. Computations with $\texttt{dimBlock} = 8$ were hardly accelerated on coarser grids, e.g., $\texttt{dimGrid} = 128$, but those with $\texttt{dimBlock} = 16$ were mostly sped up. The optimal block size is around $\texttt{dimBlock} = 64$ and a stride size around $2 \Delta h$ seems to be a feasible choice for different situations. The latter is also consistent with the finding from \cite{yang2017highly}, although non-overlapping domain decomposition is used in the current work. 



\subsection{Parallel performance}

\begin{figure}[htbp!]
\begin{center}
 \includegraphics[angle=0,width=.92\textwidth]{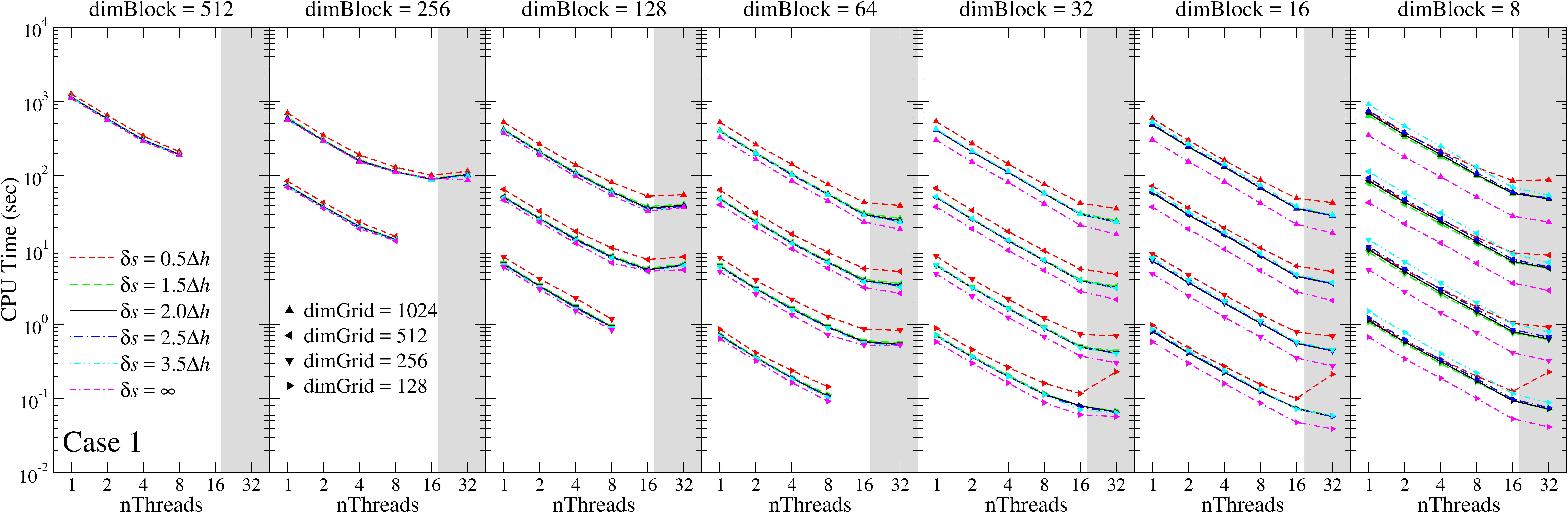}\\
\vspace{1.ex}
 \includegraphics[angle=0,width=.92\textwidth]{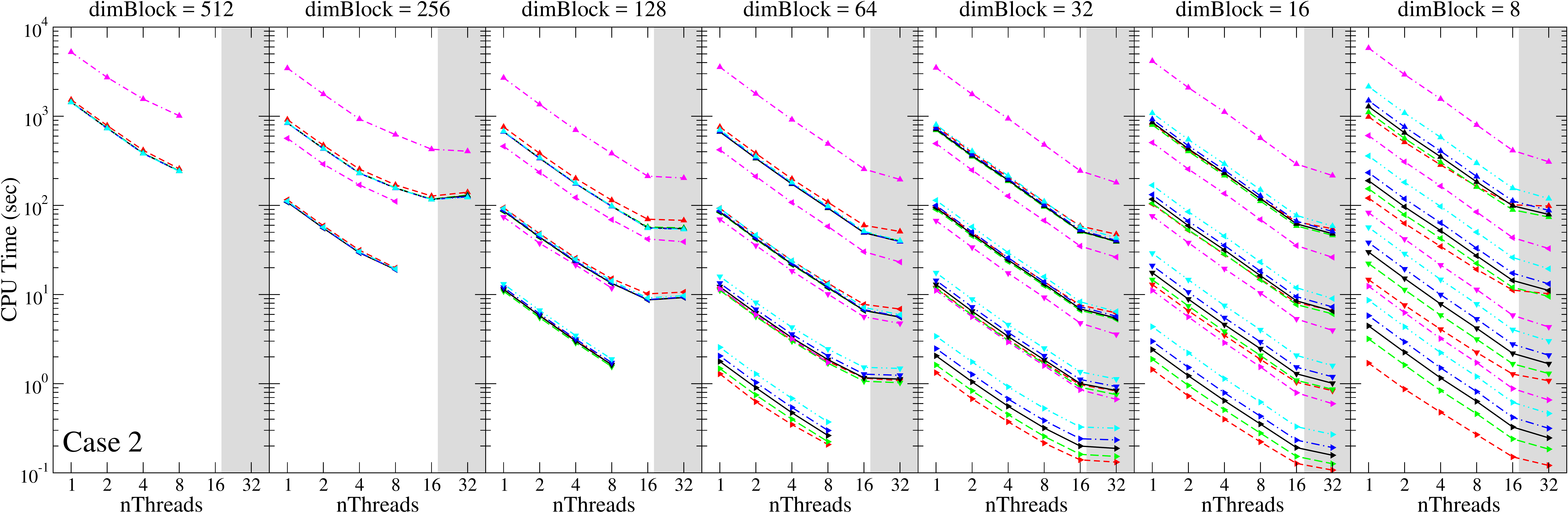}\\
\vspace{1.ex}
 \includegraphics[angle=0,width=.92\textwidth]{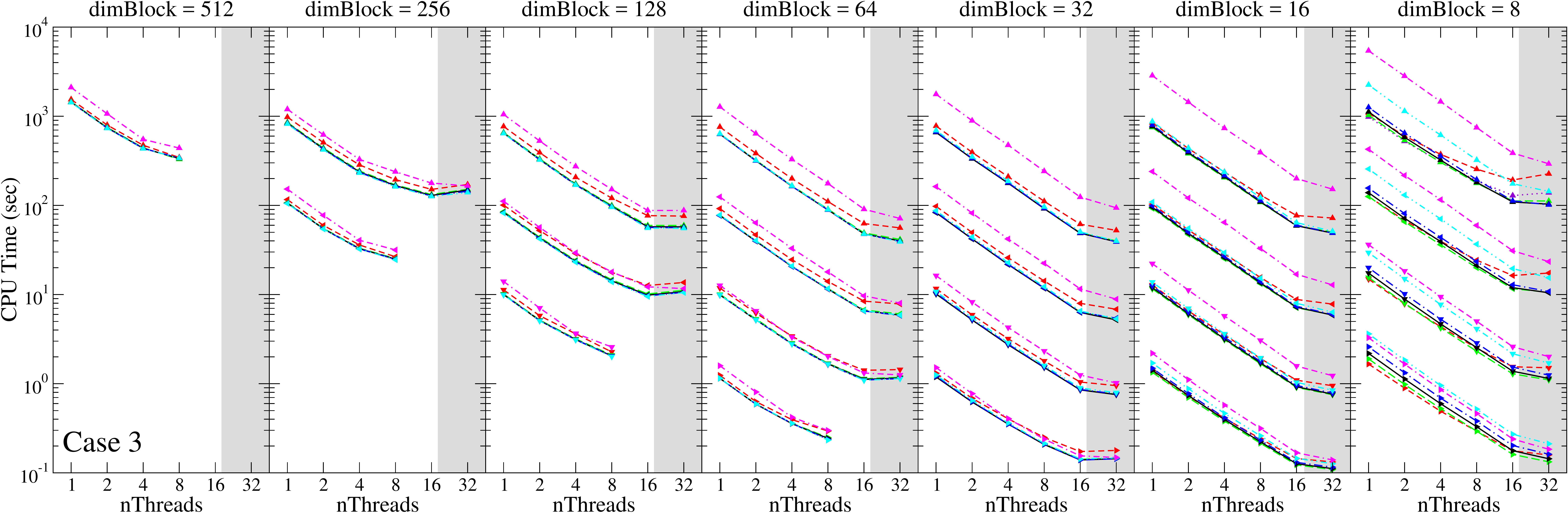}\\
\vspace{1.ex}
 \includegraphics[angle=0,width=.92\textwidth]{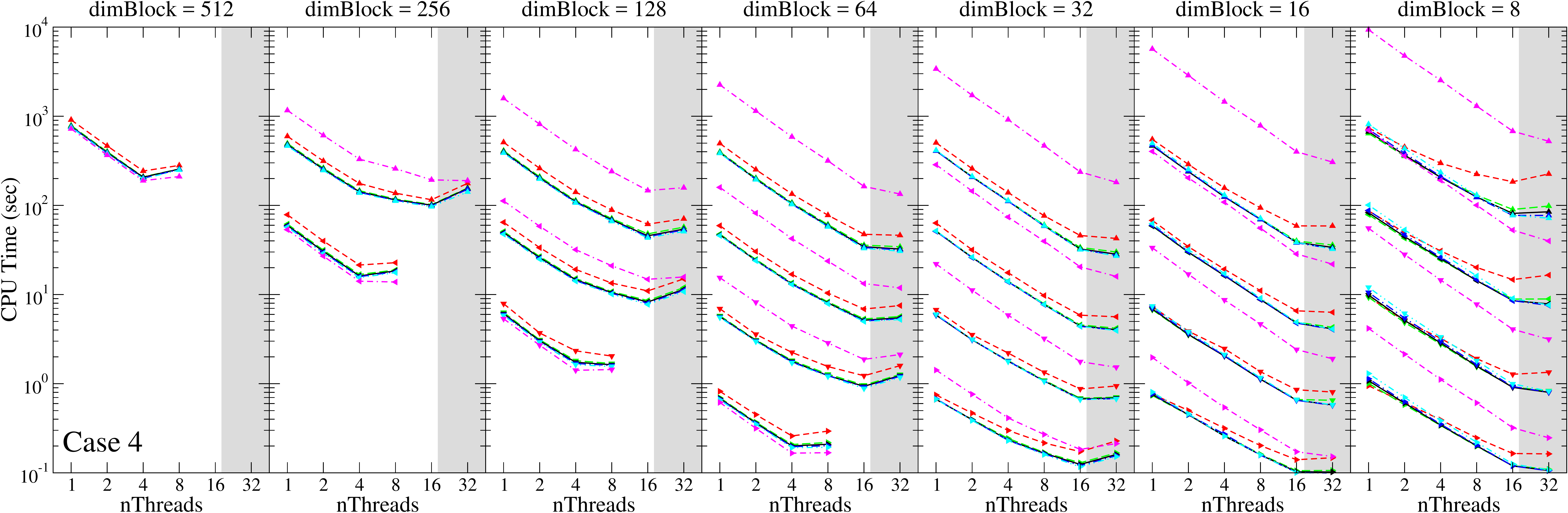}
\end{center}
 \caption{Multi-block parallel runs on AMD Ryzen Threadripper: the CPU time  $T_P$ as functions of the number of threads $\texttt{nthreads}$, the stride size $\delta s$, the block size $\texttt{dimBlock}$, and the grid size $\texttt{dimGrid}$.}
\label{fig:cputimes}
\end{figure}

\begin{figure}[htbp!]
\begin{center}
 \includegraphics[angle=0,width=.92\textwidth]{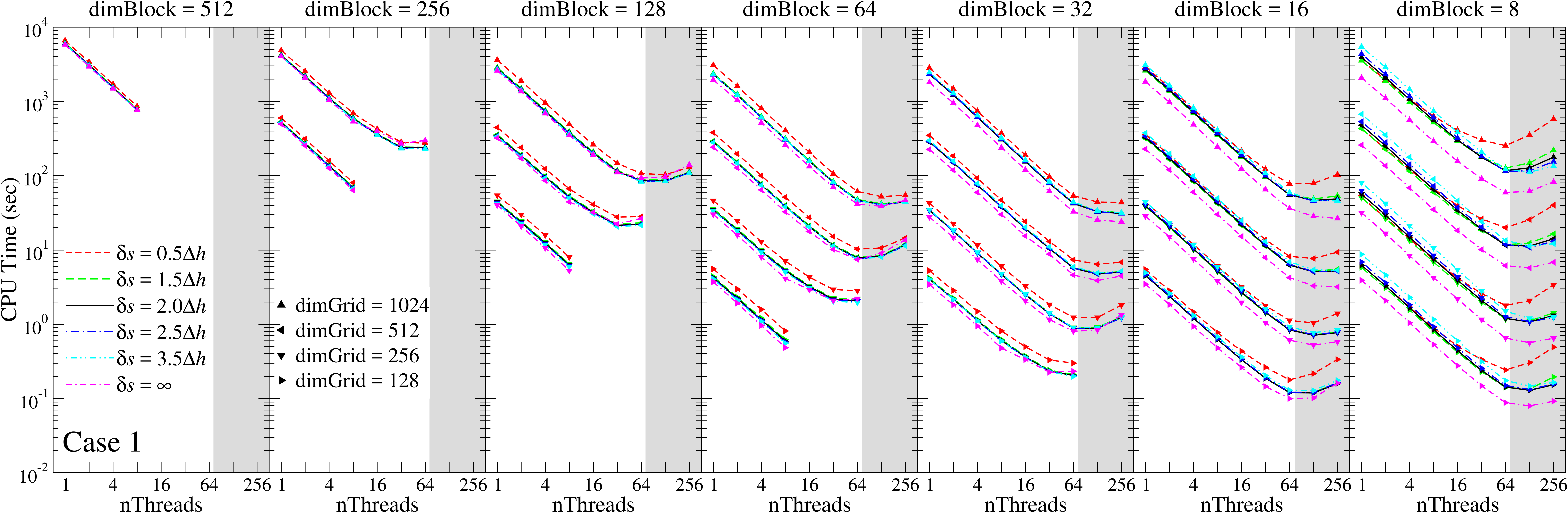}\\
\vspace{1.ex}
 \includegraphics[angle=0,width=.92\textwidth]{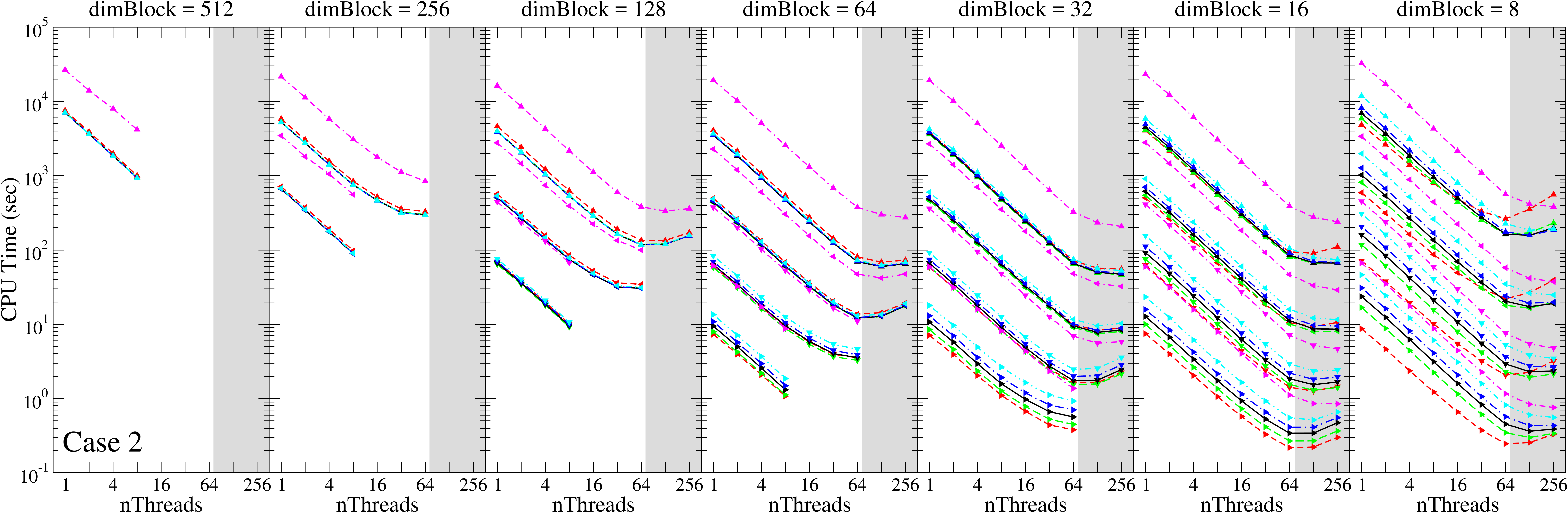}\\
\vspace{1.ex}
 \includegraphics[angle=0,width=.92\textwidth]{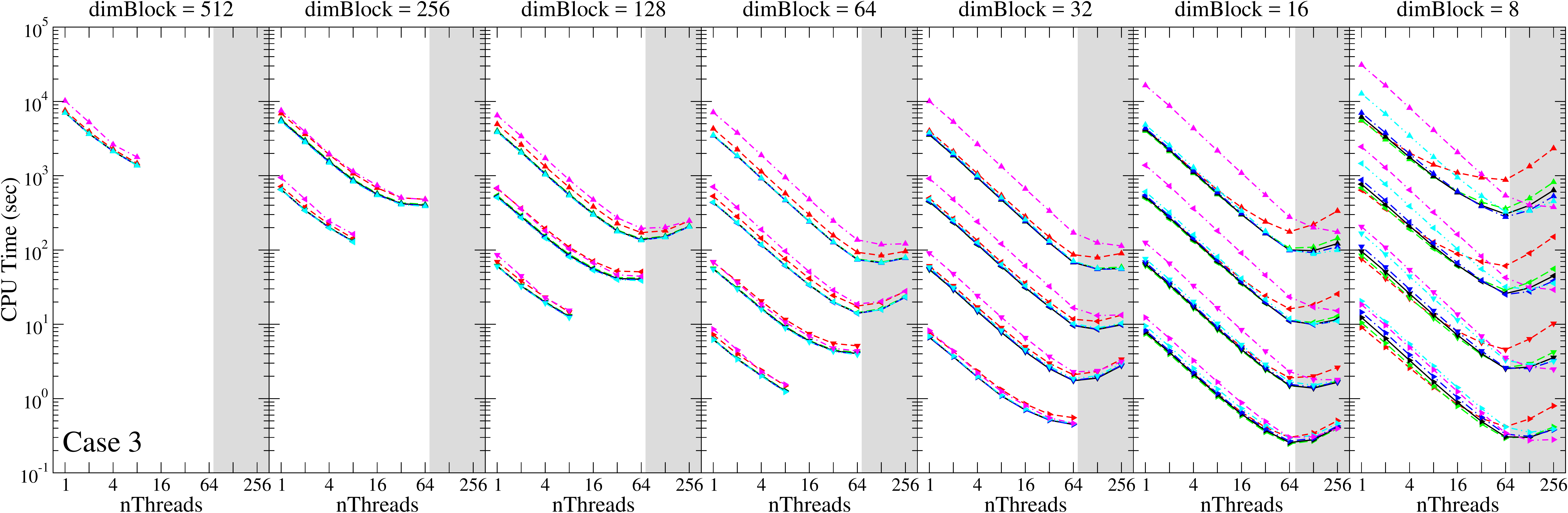}\\
\vspace{1.ex}
 \includegraphics[angle=0,width=.92\textwidth]{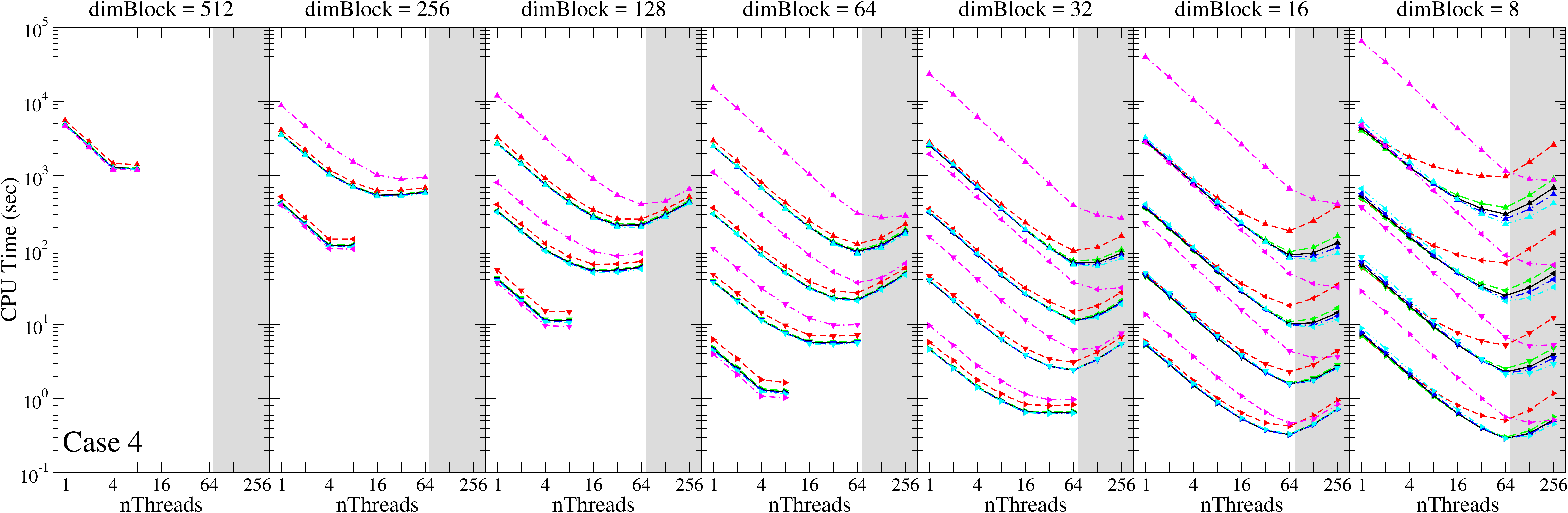}
\end{center}
 \caption{Multi-block parallel runs on Intel Xeon Phi: the CPU time  $T_P$ as functions of the number of threads $\texttt{nthreads}$, the stride size $\delta s$, the block size $\texttt{dimBlock}$, and the grid size $\texttt{dimGrid}$.}
\label{fig:cputimes_phi}
\end{figure}

Although our multi-block sequential solver considerably outperforms the baseline sequential fast marching solver as shown above, it is not the ultimate goal of current study when multi-core and many-core computers become ubiquitous in today's technical computing environment. In this part, the results from the multi-block parallel solver on two such computers are presented and discussed. The CPUs on both computers support Simultaneous Multi-Threading (SMT) technology, i.e., one physical CPU core is capable of running two or more threads simultaneously, for improved parallelization. In particular, the Xeon Phi many-core processor supports Intel's Hyper-Threading technology and is capable of running 256 threads with 64 physical CPU cores. Solving the Eikonal equation with hundreds of threads represents a shared-memory parallel computing of a significantly larger scale than previously demonstrated in the literature. To distinguish the results of parallel computations employing SMT from those using one thread per physical core, the former are marked with a grey-shaded background in the figures, i.e., the results obtained from 64 threads on the AMD Ryzen Threadripper computer and those from 128 and 256 threads on the Intel Xeon Phi computer.

Figures \ref{fig:cputimes} and \ref{fig:cputimes_phi} present the CPU time $T_P$ as functions of the number of threads $\texttt{nthreads}$, the stride size $\delta s$, the block size $\texttt{dimBlock}$, and the grid size $\texttt{dimGrid}$ on these two computers, respectively. A significant observation from these results is that all cases, regardless of the speed function, the grid size, and the choices of the two algorithm parameters, were greatly accelerated along with the increase of $\texttt{nthreads}$ from $1$ to the number of physical CPU cores on both computers. The only common exception seems to be the runs with $\texttt{nBlocks}=8$ for Case 4, which did not gain much acceleration when $\texttt{nthreads}$ increased from $4$ to $8$, apparently due to the special arrangement of the obstacles in the domain. The unfavorable effects of the obstacles in Case 4 on the parallel performance also showed up for runs with $\texttt{nBlocks}=64$ when $\texttt{nthreads}$ increased to $32$ and $64$ on the Intel Xeon Phi computer, whereas those on the AMD Ryzen Threadripper computer were not similarly affected. With  $\texttt{nthreads} = \texttt{nBlocks}$, most of these runs somehow resemble the stationary domain decomposition in \cite{yang2017highly}, in which the load imbalance among different subdomains was hardly mitigated. On the other hand, the overall parallel scalability demonstrated here is truly remarkable, as the CPU time decreased almost linearly along with the increase of the number of threads for most cases, even including $\delta s = \infty $ and $0.5 \Delta h$. 

For the parallel runs, the effects of the stride size barely show any difference from those in the sequential runs as the number of threads increases. A value around $2 \Delta h$ for $\delta s$  remains the rule-of-thumb choice for achieving a superior performance. However, there does exist a relative trend, especially for runs with $\texttt{dimBlock} \leq 32$, that the performance of a smaller $\delta s$ declines  (e.g., $\delta s = 0.5 \Delta h$) and that of a larger  $\delta s$ (e.g., $\delta s = \infty $) improves as $\texttt{nthreads} $ increases. In general, it is preferred to have the optimal values of a free parameter restricted within a stable and narrow range. The fact that the stride size here is such a free parameter confirms the robustness and applicability of our restarted narrow band approach. 

The other free parameter in our current algorithm is the block size $\texttt{dimBlock}$.  In general, the CPU time $T_P$ showed a trend of almost linear decrease along with the increase of $\texttt{nthreads}$ from $1$ to the number of physical CPU cores, except some runs with a limited number of blocks because of $\texttt{mBlocks} = 2$ (and $\texttt{mBlocks} = 4$ on the Intel Xeon Phi computer) for Case 4. In a multi-thread setting, the optimal block size seems to shift down from $\texttt{dimBlock} = 64$ for the sequential runs towards a value around $\texttt{dimBlock} = 32$. For coarser grids, the choice of $\texttt{dimBlock} = 16$ gave slightly better results than $\texttt{dimBlock} = 32$ for some runs. On the other hand,  $\texttt{dimBlock} = 64$ provided the best performance on finer grids for some cases on the AMD Ryzen Threadripper computer. Therefore, similar to the situation of the stride size discussed above, the block size $\texttt{dimBlock}$ in our block-based algorithm has a stable narrow range for achieving the best performance. The rule-of-thumb choice is a value around $\texttt{dimBlock} = 32$, and it can be shifted up for a very fine grid or down for a coarse one to obtain a comparable or moderately improved performance. Another preferred feature that can be observed in the results is the apparent independence between the block size and the stride size. As discussed above, some results with $\delta s = 0.5 \Delta h$ showed a weak reliance on the block size for runs with $\texttt{dimBlock} = 8$ and $16$. But again, a very small stride size such as $\delta s = 0.5 \Delta h$ is not recommended to be used in practical applications of the restarted narrow band approach anyway. 

The results also demonstrated the effects of SMT on the parallel performance. It is evident that on both computers SMT barely offered any improvements when $\texttt{dimBlock} \geq 128$ except a few runs on the finest grids. With smaller block sizes on the AMD Ryzen Threadripper computer, however, the CPU time continued to decrease when the number of threads on each physical core increased from one to two. The SMT acceleration on finer grids was actually rather prominent, which illustrates the unambiguous benefits of SMT on this computer with our algorithm. On the other hand, running the multi-thread solver with SMT (or hyper-threading) activated on the Intel Xeon Phi computer showed disappointing, mostly negative effects on the parallel performance. Except for a few runs (e.g., $\texttt{dimBlock} = 32$ on grid $\texttt{dimGrid} = 1024$ for Case 1 and Case 2), four threads on each physical CPU core, for a total of $256$ threads, generally required more time to complete the computations compared with the other scenarios (one or two threads per core). The runs with $128$ threads through two threads per core showed moderate gains on finer grids in most cases except Case 4. Surprisingly, the runs with $\delta s = \infty $ seem to benefit from hyper-threading on this computer, especially for $\texttt{dimBlock} = 32$ and $16$. It should be noted that, apart from differences between the two computers, the load imbalance also played a larger role here for the less encouraging SMT results compared with those on the AMD Ryzen Threadripper computer: e.g., for $\texttt{mBlocks} = 8$ the total number of blocks would be $512$, then with $\texttt{nthreads} = 128$ or $256$ each thread would be handling only $4$ or merely $2$ blocks! And this is the scenario for $\texttt{dimBlock} = 128$ on grid $\texttt{dimGrid} = 1024$,  $\texttt{dimBlock} = 64$ on grid $\texttt{dimGrid} = 512$,  $\texttt{dimBlock} = 32$ on grid $\texttt{dimGrid} = 256$, or $\texttt{dimBlock} = 16$ on grid $\texttt{dimGrid} = 128$.

\begin{figure}[htbp!]
\begin{center}
 \includegraphics[angle=0,width=.92\textwidth]{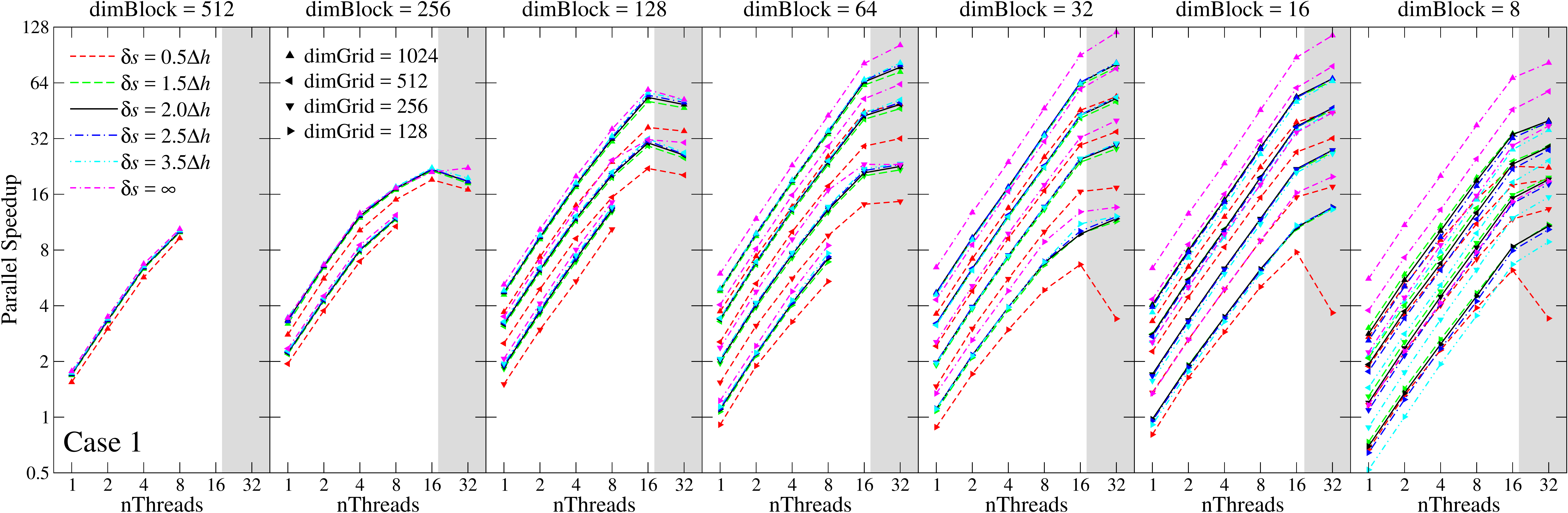}\\
\vspace{1.ex}
 \includegraphics[angle=0,width=.92\textwidth]{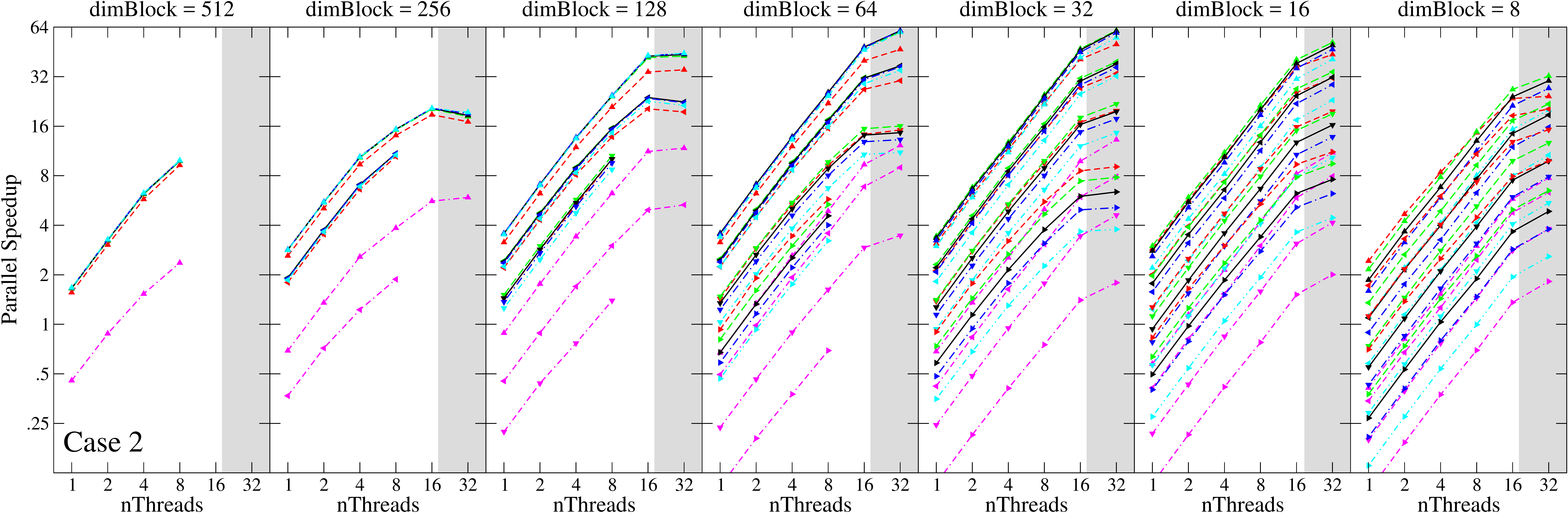}\\
\vspace{1.ex}
 \includegraphics[angle=0,width=.92\textwidth]{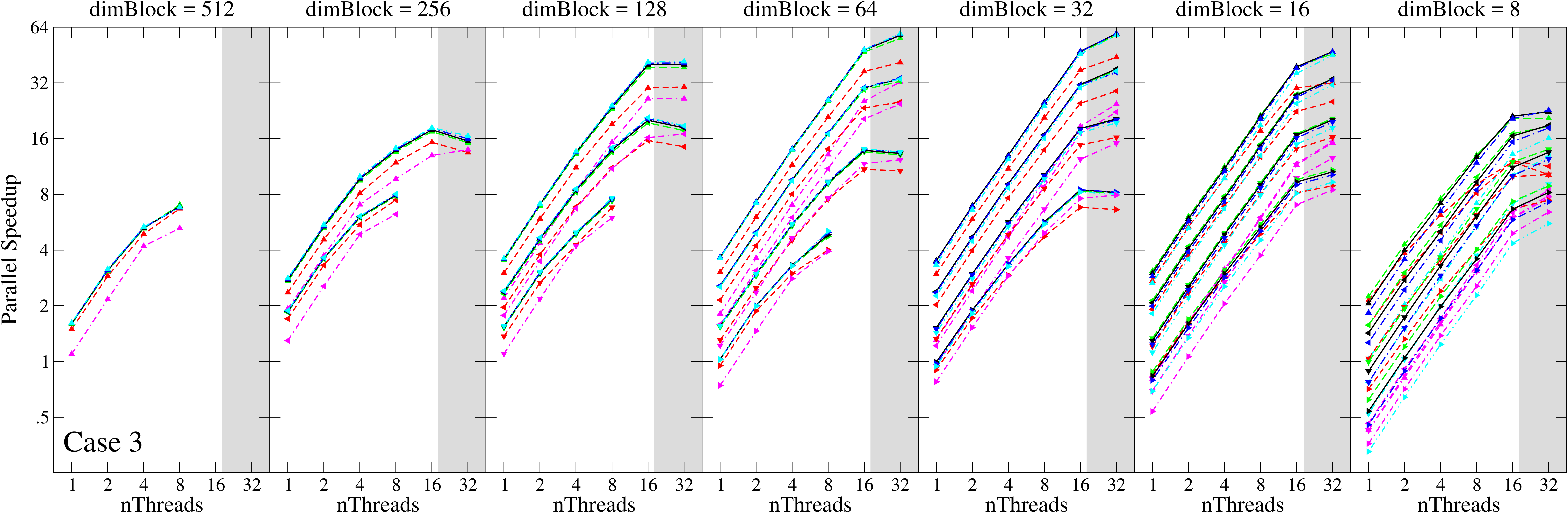}\\
\vspace{1.ex}
 \includegraphics[angle=0,width=.92\textwidth]{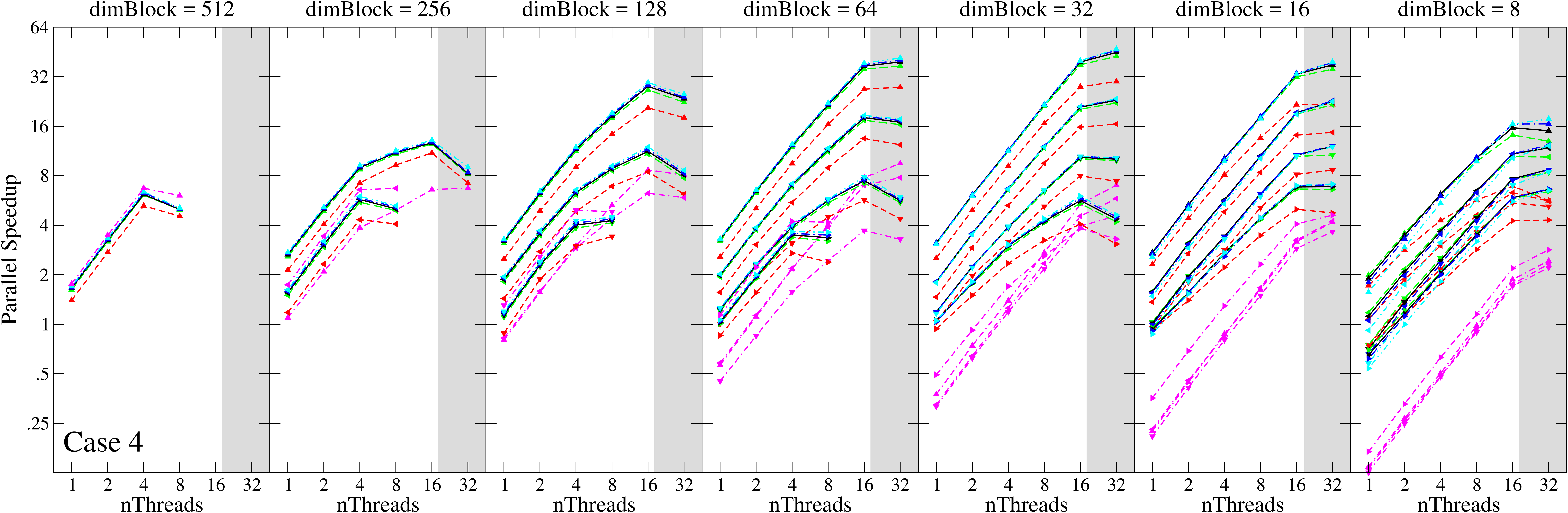}
\end{center}
 \caption{Multi-block parallel runs on AMD Ryzen Threadripper: the parallel speedup as functions of the number of threads $\texttt{nthreads}$, the stride size $\delta s$, the block size $\texttt{dimBlock}$, and the grid size $\texttt{dimGrid}$.}
\label{fig:speedups}
\end{figure}

\begin{figure}[htbp!]
\begin{center}
 \includegraphics[angle=0,width=.92\textwidth]{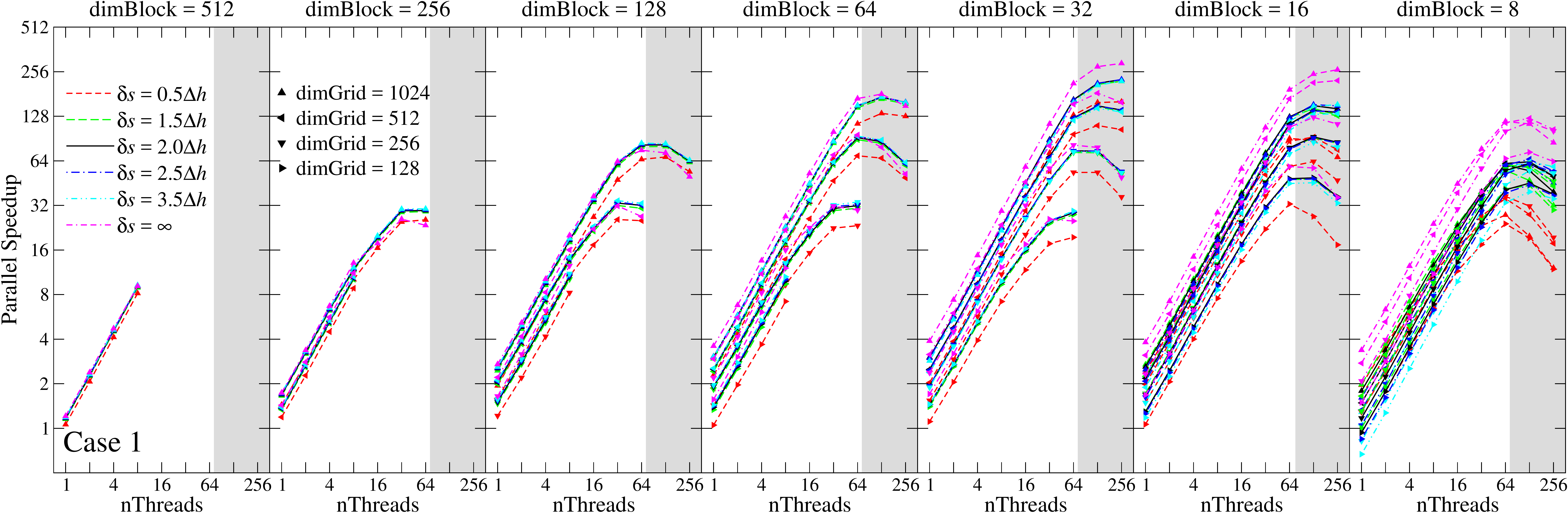}\\
\vspace{1.ex}
 \includegraphics[angle=0,width=.92\textwidth]{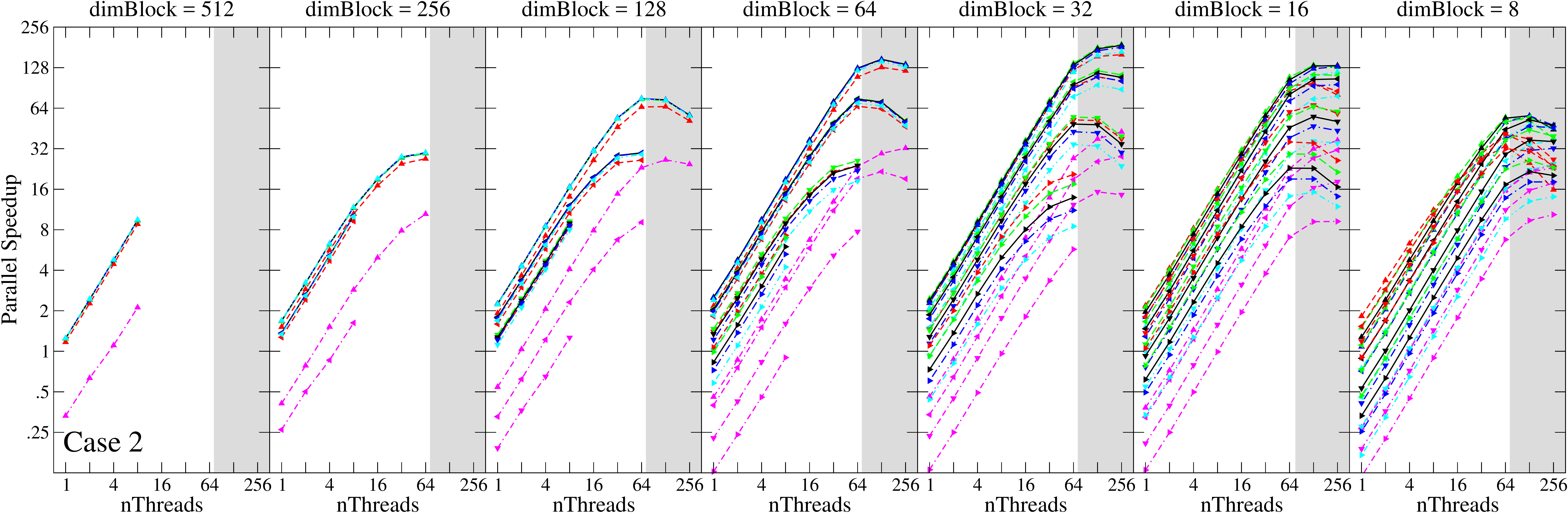}\\
\vspace{1.ex}
 \includegraphics[angle=0,width=.92\textwidth]{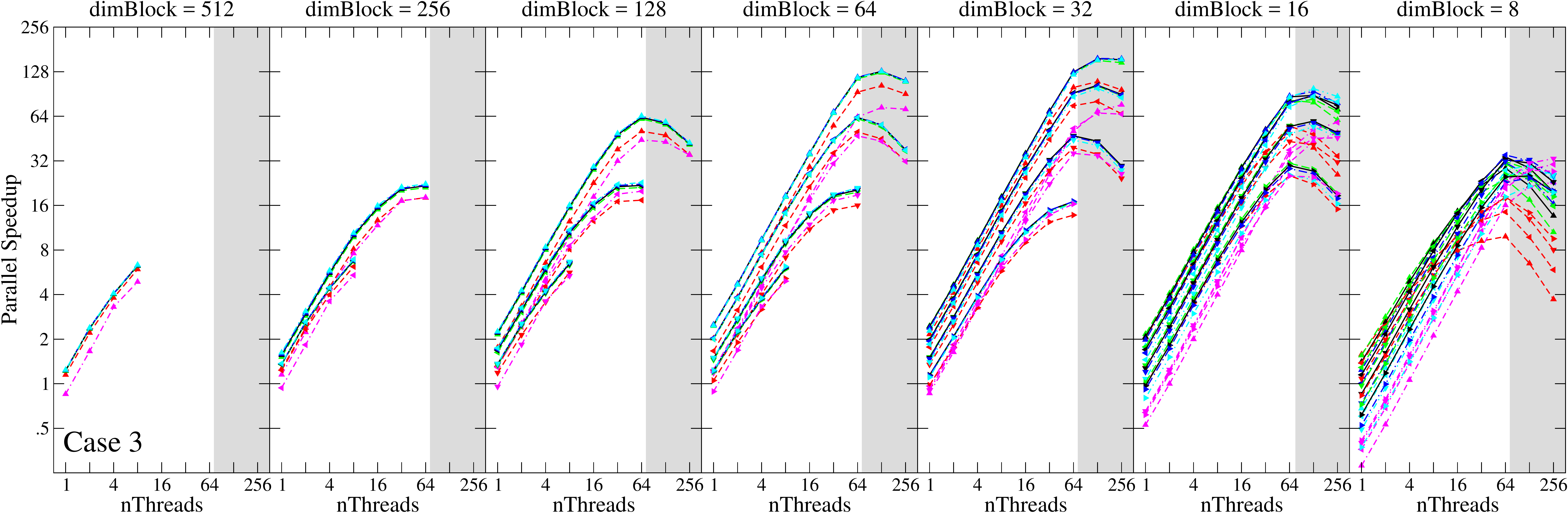}\\
\vspace{1.ex}
 \includegraphics[angle=0,width=.92\textwidth]{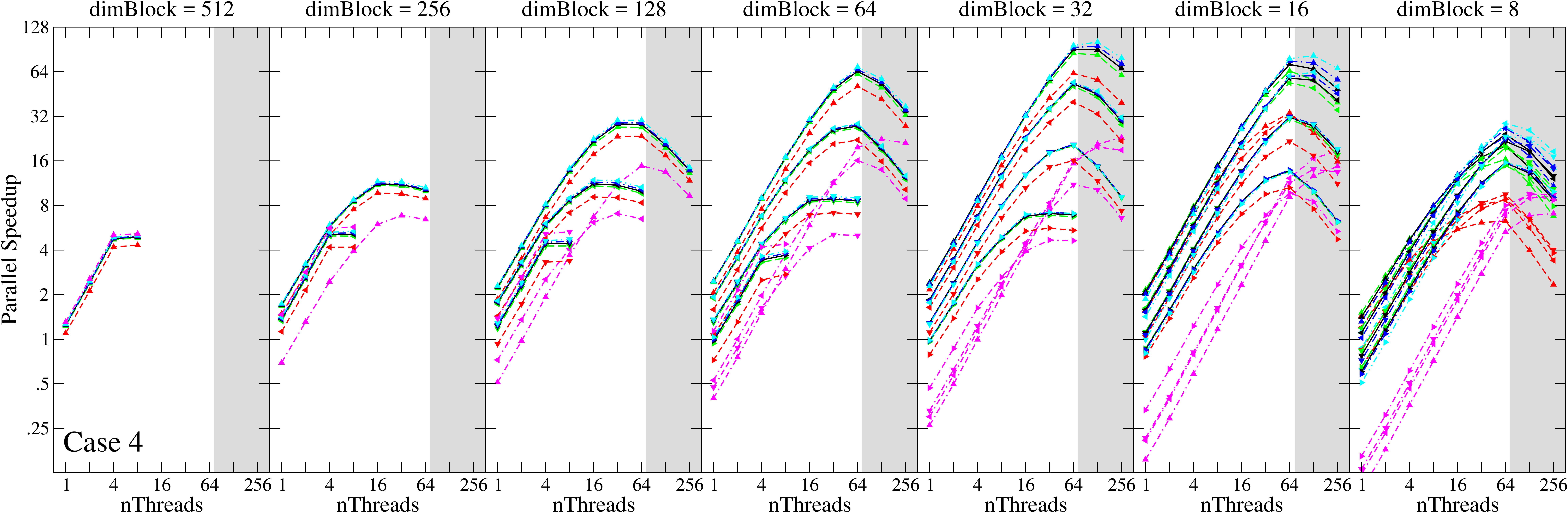}
\end{center}
 \caption{Multi-block parallel runs on Intel Xeon Phi: the parallel speedups as function of the number of threads $\texttt{nthreads}$, the stride size $\delta s$, the block size $\texttt{dimBlock}$, and the grid size $\texttt{dimGrid}$.}
\label{fig:speedups_phi}
\end{figure}

Figures \ref{fig:speedups} and \ref{fig:speedups_phi} show the corresponding parallel speedups on both computers. Here the absolute speedup is used, which is defined as $S = {T _P}/{T _B}$ with $T_B$ the CPU time obtained from the single-block sequential solver. It is also possible to use the relative speedup defined using the CPU time in this part with $\texttt{nthreads} = 1$ as the reference, but in that case it could not be a single value as the CPU time of the multi-block solver depends on the choices of $\texttt{dimBlock}$ and $\delta s$. Since the multi-block solvers can be several times faster than the single-block solver as shown in the previous part, it is interesting to notice that most curves in these two figures start from points much higher than one at $\texttt{nthreads} = 1$. Moreover, because the CPU time and the number of threads are almost inversely proportional starting from $\texttt{nthreads} = 1$ in most cases, the combined effect is that the maximum speedup can be much higher than the number of threads. For instance, on the AMD Ryzen Threadripper computer, the maximum speedup with $32$ threads on $16$ cores is more than $80$ for Case 1 (up to $120$ with $\delta s = \infty $), more than $60$ for Case 2, more than $59$ for Case 3, and more than $47$ for Case 4, all on the finest grid. On the Intel Xeon Phi computer, the maximum speedup with up to $256$ threads on $64$ cores is more than $220$ (up to $291$ with $\delta s = \infty $) for Case 1, up to $190$ for Case 2, up to $157$ for Case 3, and up to $103$ for Case 4, also all on the finest grid. These data showing the super-linear speedup phenomena are highly remarkable for a parallel algorithm of the fast marching method, which was long considered inherently sequential. Furthermore, the parallel performance on coarser grids is also very impressive, with a maximum speedup surpassing or close to the number of cores in most tests. 

\begin{figure}[htbp!]
\begin{center}
 \includegraphics[angle=0,width=.92\textwidth]{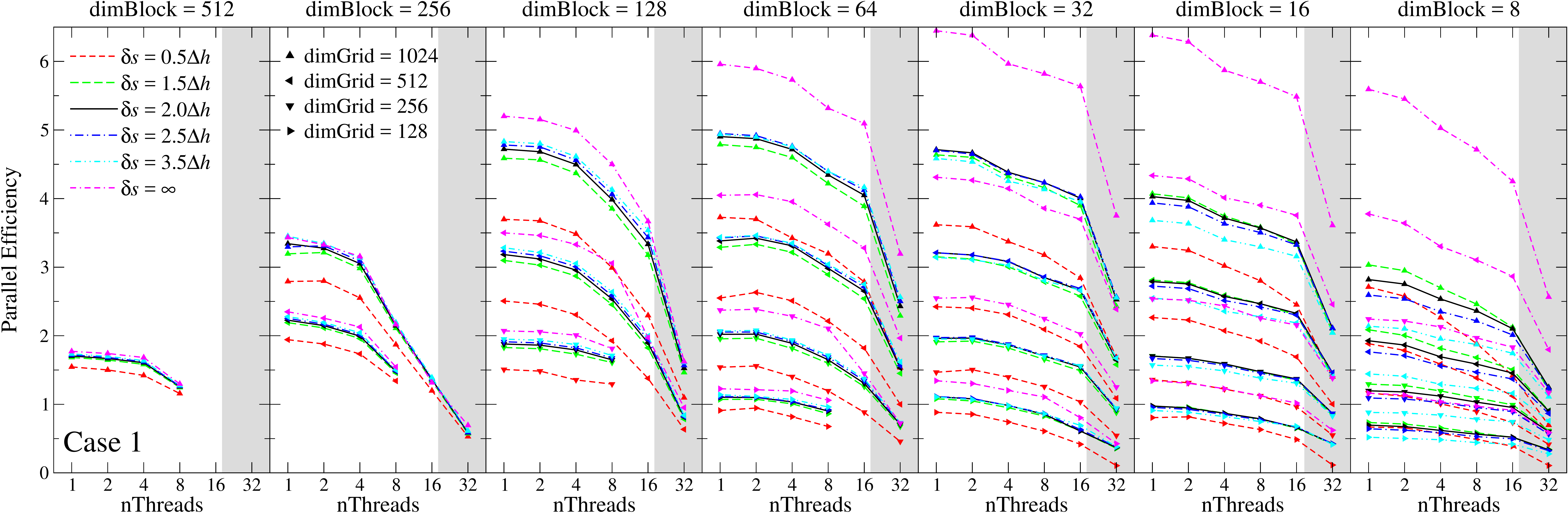}\\
\vspace{1.ex}
 \includegraphics[angle=0,width=.92\textwidth]{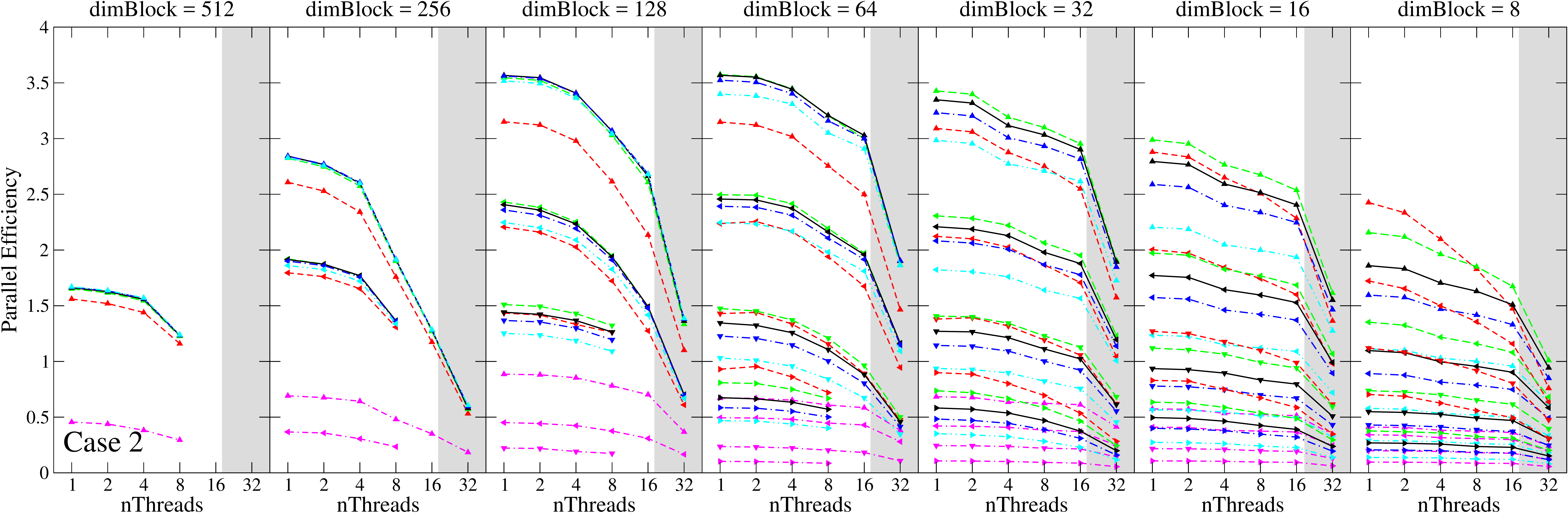}\\
\vspace{1.ex}
 \includegraphics[angle=0,width=.92\textwidth]{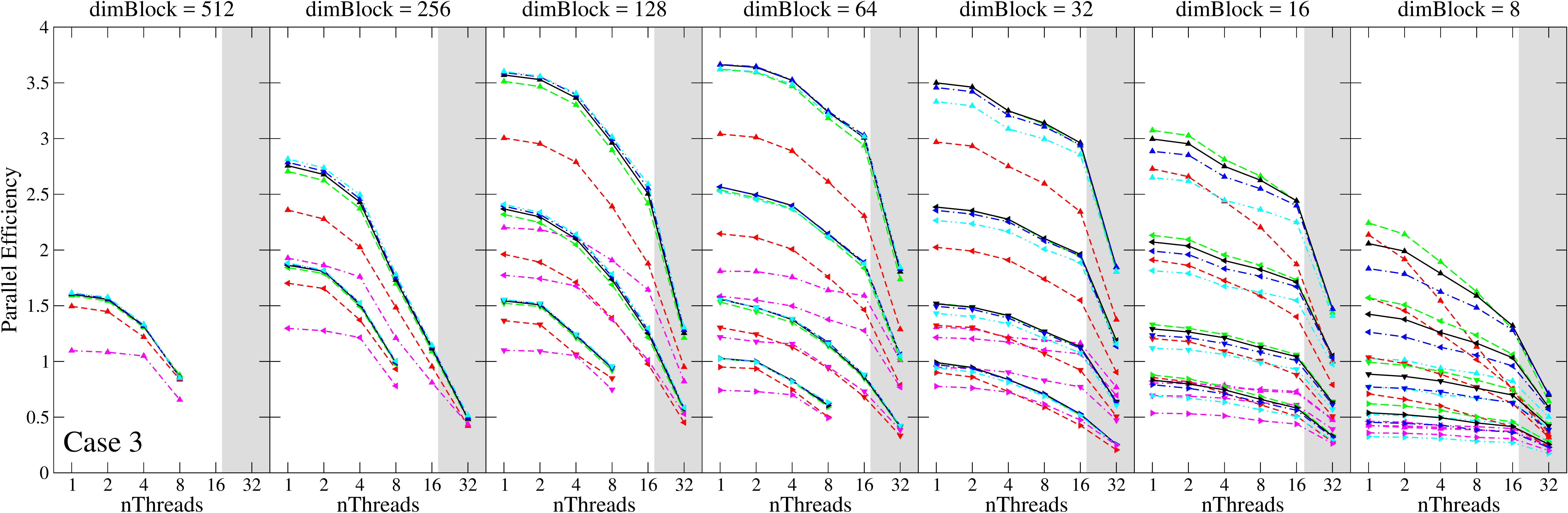}\\
\vspace{1.ex}
 \includegraphics[angle=0,width=.92\textwidth]{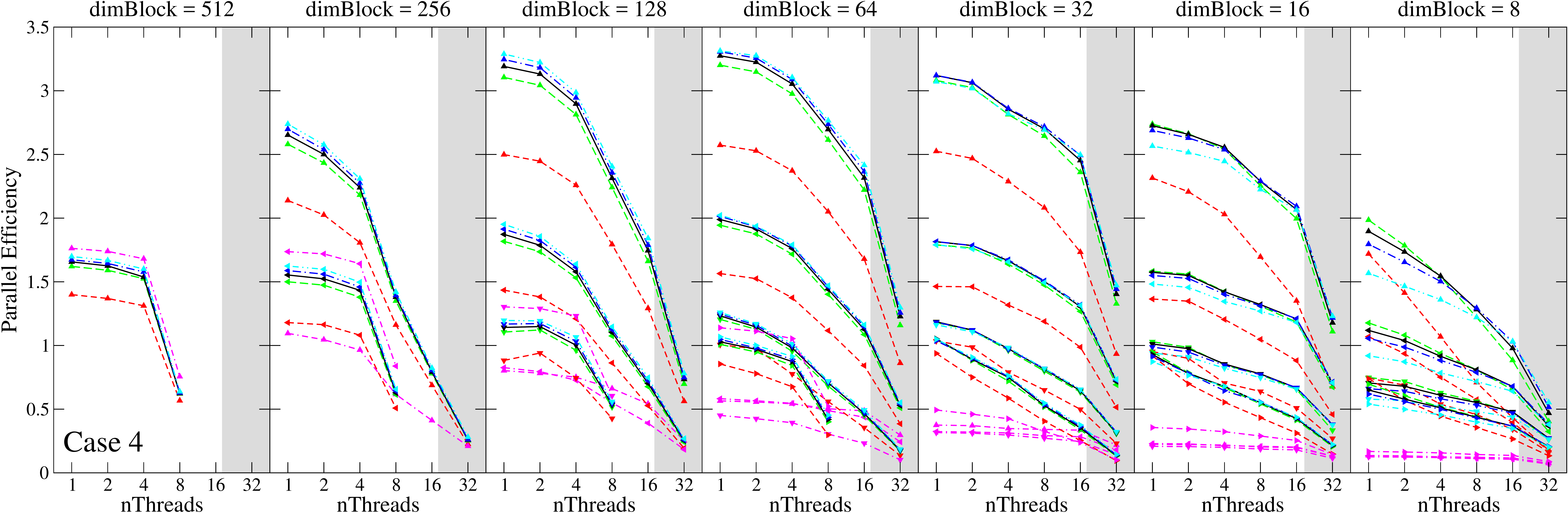}
\end{center}
 \caption{Multi-block parallel runs on AMD Ryzen Threadripper: the parallel efficiency as functions of the number of threads $\texttt{nthreads}$, the stride size $\delta s$, the block size $\texttt{dimBlock}$, and the grid size $\texttt{dimGrid}$.}
\label{fig:efficiencies}
\end{figure}

\begin{figure}[htbp!]
\begin{center}
 \includegraphics[angle=0,width=.92\textwidth]{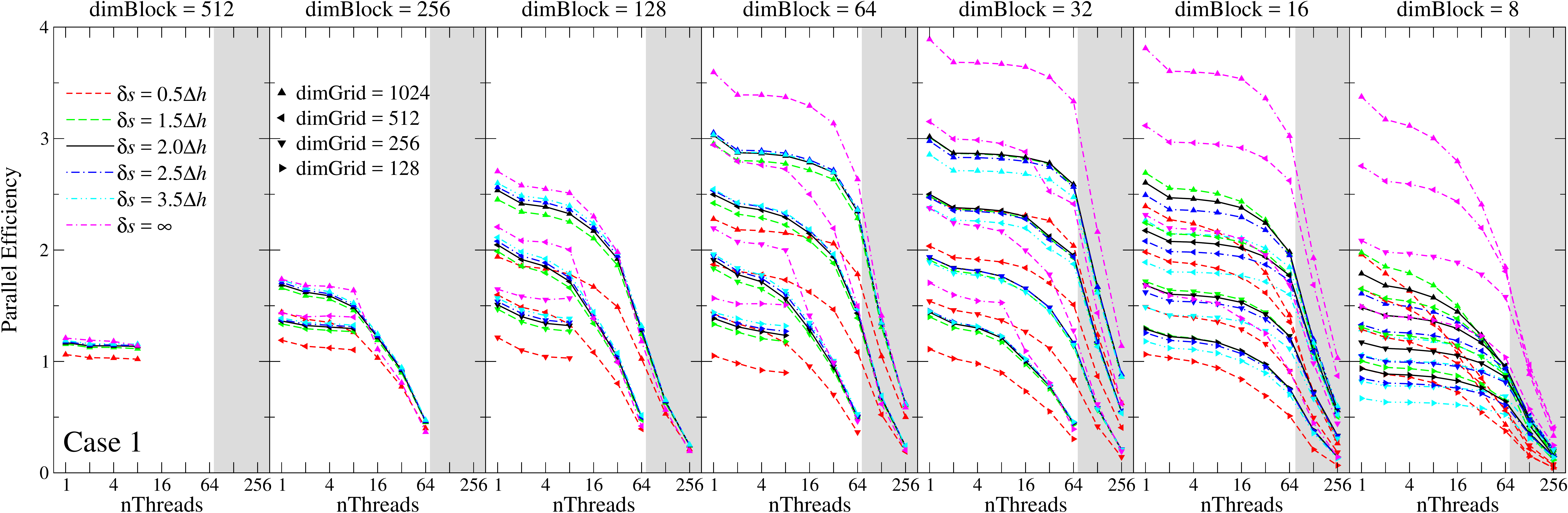}\\
\vspace{1.ex}
 \includegraphics[angle=0,width=.92\textwidth]{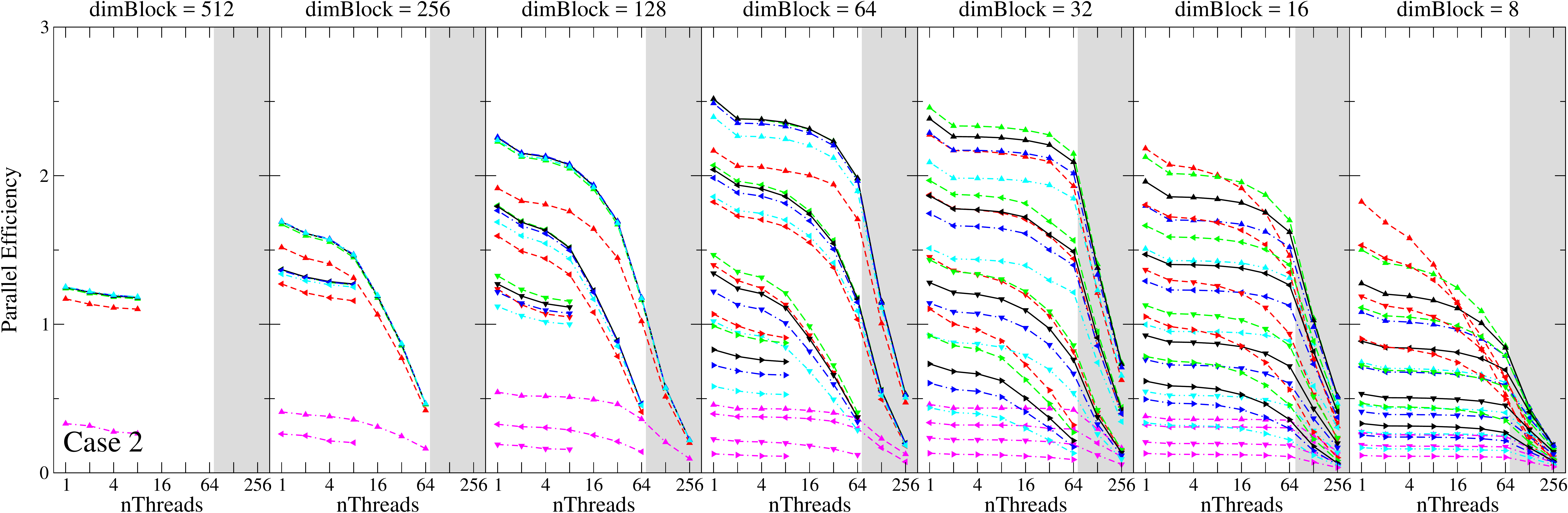}\\
\vspace{1.ex}
 \includegraphics[angle=0,width=.92\textwidth]{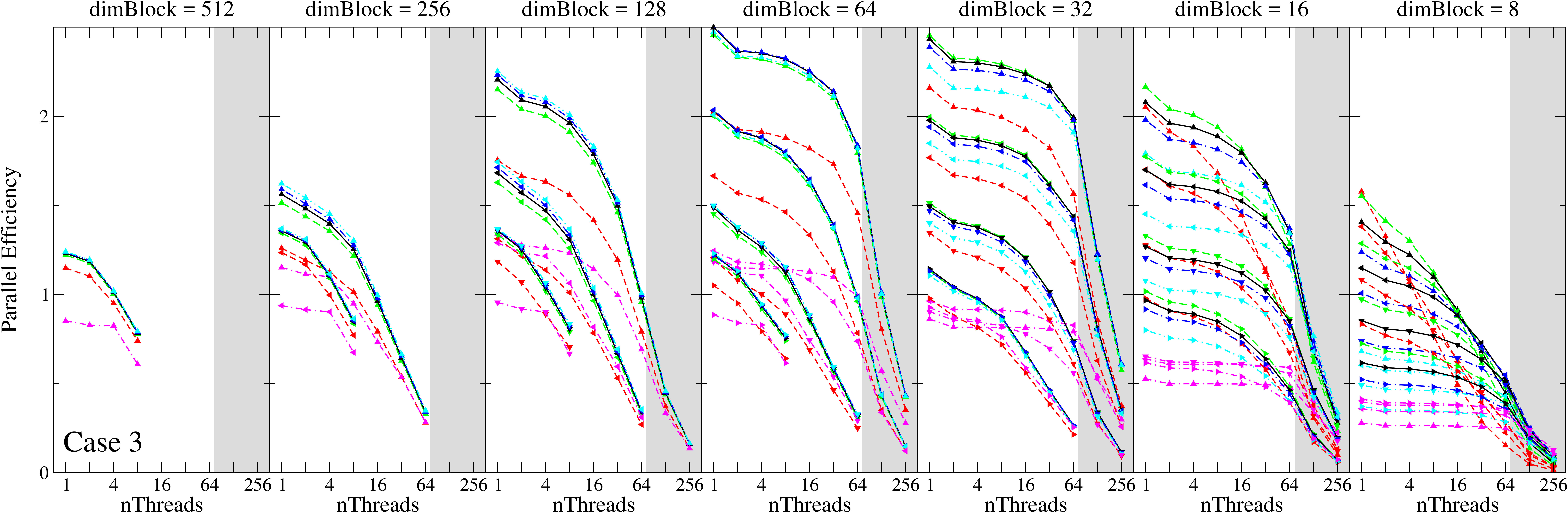}\\
\vspace{1.ex}
 \includegraphics[angle=0,width=.92\textwidth]{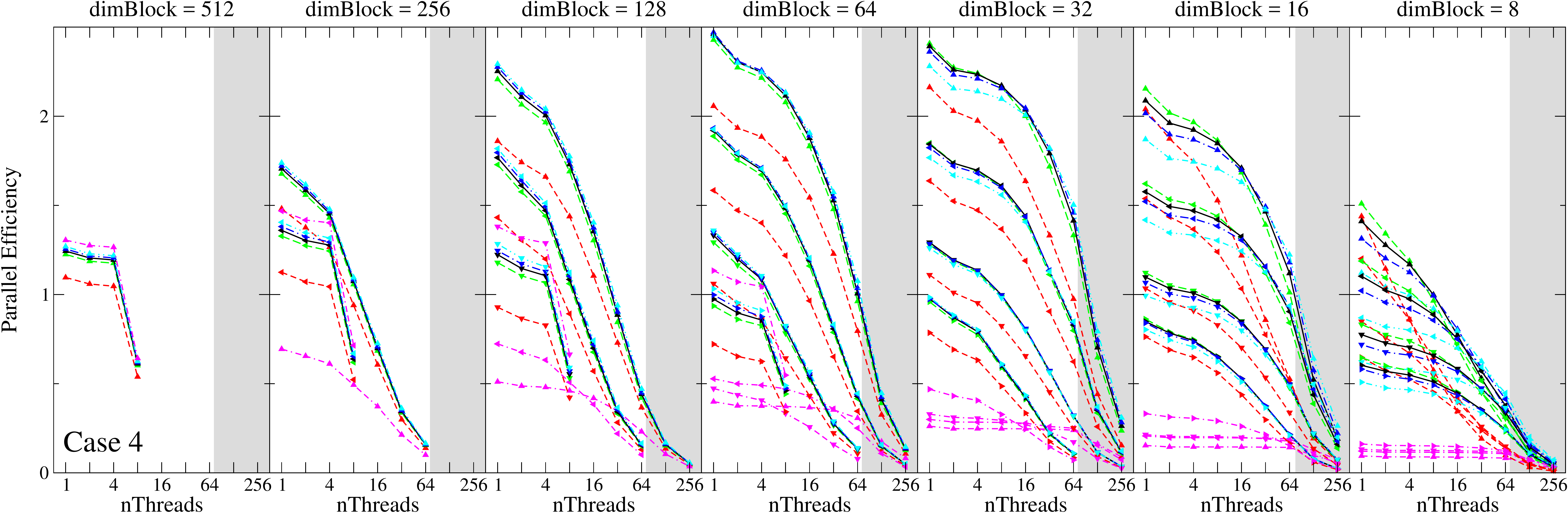}
\end{center}
 \caption{Multi-block parallel runs on Intel Xeon Phi: the parallel efficiency as functions of the number of threads $\texttt{nthreads}$, the stride size $\delta s$, the block size $\texttt{dimBlock}$, and the grid size $\texttt{dimGrid}$.}
\label{fig:efficiencies_phi}
\end{figure}

The corresponding parallel efficiency, which is defined as $E = S/\texttt{nthreads} $, is shown in Figures \ref{fig:efficiencies} and \ref{fig:efficiencies_phi} for tests on both computers. The comprehensive super-linear speedup phenomena of our parallel algorithm, shows as parallel efficiencies higher than $E=1$ in the figures, become even easier to recognize than those in the speedup plots above.  It is also evident in the figures that, in terms of free parameters, $\texttt{dimBlock} = 32$ and $\delta s = 2 \Delta h$ are the overall optima for giving consistently high parallel efficiencies for vastly different speed functions and grid sizes. In addition, we can barely recognize any interdependence between the block size and the stride size, if neglecting the extreme cases $\delta s = \infty $ and $0.5 \Delta h$.

\section{Discussions}

Our algorithm generally shows a lower parallel efficiency on the Intel Xeon Phi computer than that on the AMD Ryzen Threadripper computer. The AMD Ryzen Threadripper 1950X processor has 8 MB L2 cache (512 KB per core) and 32 MB L3 cache, whereas the Intel Xeon Phi 7210 has 32 MB L2 cache (two cores share 1 MB L2 cache). One particular feature of the Intel Xeon Phi processor is that it has no L3 cache, but has 16GB on-package high-bandwidth memory called MultiChannel Dynamic Random Access Memory (MCDRAM), which can provide a total memory bandwidth of five times of the off-package DDR (Double Data Rate) memory (up to 450 GB/sec vs. up to 90 GB/sec), but with a similar idle latency (approximately 150 ns vs. approximately 125 ns). Different programming modes are available for the MCDRAM and DDR memory. Here we chose the default modes, i.e., the Cache Mode for the former (100\% MCDRAM as cache) and the quadrant cluster mode for the latter. It is well-known that a binary heap is not a cache friendly data structure due to the difficulties of maintaining the locality of references in heap operations. In this work, a simple 3D array was used to store the function value, status tag, and heap position of each grid point in each subdomain, thus no particular optimizations were implemented to improve heap performance. Apparently the cache miss penalties of using MCDRAM (although in Cache Model) were much higher than those of a real L3 cache due to the much higher latency of the former. This also explains why using hyper-threading with two or four threads per core hardly boosted the parallel performance as shown earlier. Nonetheless, the superior performance of our algorithm was still perfectly demonstrated on this 64-core computer by the extraordinary parallel speedup of two orders of magnitude for widely varying speed functions.

A very attractive characteristic of the fast marching method is its robustness in dealing with complex speed functions, and this feature is retained in our current algorithm. Of course, the CPU time still increases as the range of variations in the speed function increases. But the variations of CPU time generally are limited within a very narrow range, as shown in the results of different speed functions on the same grid. On the other hand, in the last case, the front propagates through narrow curved passages between multiple nonpenetrable barriers ($F = 0$), which could present a major challenge to many iterative methods in terms of the overall efficiency.

Another important feature of our new algorithm is its resemblance to the original sequential fast marching method. In the block-based algorithm, the data structures for each block exactly follow the original single-block sequential version, except a few global arrays for storing the activation statuses of each block in the marching and exchange steps. It should be noted that the block activation mechanisms do create two lists of blocks in Algorithm \ref{alg:pfmm}. But unlike the ordered lists in some other approaches \cite{WeberDBBK08,ChaconV2015}, the elements in the lists do not possess any type of priority at all in our algorithm and they are handled by different threads in orders defined by the OpenMP loop scheduling mechanism. The main change is the addition of Algorithm \ref{alg:collect}, which actually shares the same major operations with Algorithm \ref{alg:update}.

The restarted narrow band approach in the current study still closely follows the original one proposed in \cite{yang2017highly}, although with several simplifications. Depending on the speed function and the choice of grid size, domain decomposition configuration, and stride size, there are a various amount of refreshing computations in a multi-block run that are extra to a single-block run. In the multi-block sequential results examined earlier, since only one CPU core was involved, the CPU time differences among blocks (i.e., load imbalance for parallel computing) do not affect the overall performance. Therefore, the main contributor to the remarkable acceleration shown there is the reduced heap size, as each block has its own heap. In terms of algorithm complexity, the worst-case scenario is  $O(N \log N)$  for a total of $N$ grid points. Now in a multi-block setting with a total of $\texttt{nBlocks}$,  we have $\texttt{nBlocks} \times \left (\frac{N}{\texttt{nBlocks}} \log \frac{N}{\texttt{nBlocks}} \right ) = N \log \frac{N}{\texttt{nBlocks}}$. The savings of heap operations can be substantial for a large $\texttt{nblocks}$. Of course, because each block is padded with ghost cells, the memory overhead extra to a single-block run grows with $\texttt{nblocks}$ too. Nonetheless, the case with many short heaps apparently performs better than the one with one single long heap in most situations. It should be noted that sometimes the overhead of refreshing computations due to a complex speed function can be significant, as shown in the multi-block sequential runs on grid $\texttt{dimGrid} = 128$ for Case 2. The effectiveness of our restarted narrow band strategy in reducing excessive refreshing computations can be easily grasped by comparing the results from a stride size around $\delta s = 2 \Delta h$ and those of $\delta s = \infty$. Of course, sometimes $\delta s = \infty$ may give a better performance, when the interdependence between blocks is weak or simple, as shown in some results for Case 1 ($F=1$). But when a complex speed function results in a strong interdependence between blocks, such as in Case 2, the importance of the restarted narrow band approach becomes evident. 

In the multi-thread runs, the tests with $\texttt{nthreads}  = \texttt{nblocks}$ encountered the condition similar to that in \cite{yang2017highly} with a one-to-one relationship between threads and blocks. In such circumstances, computations on some threads may become intermittent and the load imbalance among threads cannot be avoided except for a perfectly divided load like $\texttt{nblocks} = 8$ for Case 1. Of course, $\texttt{nthreads}  = \texttt{nblocks}$ is a unique scenario, but it is conceivable that the block activation mechanisms introduced in this study won't work as expected if the number of blocks is equal or close to the number of threads. In general, our current approach relies on an adequately decomposed domain to ensure $\texttt{nblocks} \gg \texttt{nthreads}$, such that the block activation mechanisms can kick in meaningfully to skip blocks not engaged with the front. This is verified by the comprehensive superior performance on finer grids with the same block size. On the other hand, as $\texttt{mblocks}$ grows, the memory overhead for padding ghost cells also grows fast, as shown by the prevailing performance descent shown in the results when the block size was reduced to $\texttt{dimBlock} = 8$. Fortunately, our results have shown that the choice of $\texttt{dimBlock} = 32$ reasonably balances the trade-off between load imbalance and memory overhead. It is interesting to note that as the number of threads increases, the optimal block size seems to shift from $64$ toward $32$. This could possibly be explained by the smaller heap sizes and more balanced loads with $\texttt{dimBlock} = 32$ in spite of the memory overhead increase.

\section{Conclusions}\label{sec:conclusions}

A block-based fast marching method with superior sequential and parallel peformance has been developed in this paper. The computational domain is decomposed into non-overlapping blocks that are padded with ghost cells. The original sequential fast marching algorithm is essentially run on each block. An exchanging step is added to update ghost cells using data from neighboring blocks, such that earlier computations can possibly be refreshed with any incoming front characteristics. Both the marching and exchanging steps are placed in an infinite loop with an exit criterion similar to that of the original sequential fast marching method, but to be satisfied on all blocks. What distinguishes our block-based algorithm from a na\"{i}ve low-performance implementation is the restarted narrow band approach, first introduced in \cite{yang2017highly} and considerably simplified in this paper, with which the global narrow band bound is replaced by an incremental one, increasing by a given stride in each restart. This strategy can greatly reduce excessive refreshing computations through timely synchronized exchanges. In addition, depending on its value, an accepted point can be repeatedly updated later, and it may be excluded from the stencil as an upwind point for updating a neighbor. Especially, a grid point initialized as a boundary condition is separately labeled, such that it can be inserted into the heap and treated like other points, but both its value and status won't be possibly changed like them. These minor modifications are fully consistent with the original sequential algorithm. In particular, simple block activation mechanisms are introduced for both the marching and exchanging steps, so a block will only be included for running one or both steps when it is involved with the propagating front. Therefore, such a multi-block algorithm mainly consists of two loops for carrying out the marching and exchanging steps on two groups of blocks, which can be easily parallelized using OpenMP on a shared-memory multi-core and/or many-core computer. Surprisingly, the whole algorithm does not involve any data race conditions and is free of OpenMP critical construct or lock synchronization. It is also interesting to note that two advantageous features of the fast marching method are perfectly retained, i.e., the monotonic increasing order of the solution and the narrow band formulation, which makes our new algorithm an ideal benchmark parallel implementation of the original sequential fast marching method. Detailed pseudo-codes are provided to illustrate the modification procedure from the single-block sequential algorithm to the multi-block one in a step-by-step manner.

Four point source problems with various speed functions typical in practical applications have been systematically tested on four grids ranging from $128^3$ to $1024^3$ points on two desktop computers using up to $256$ threads. Six stride sizes including the extreme option of $\delta s = \infty$ were examined and up to eight block sizes were evaluated. The relative differences in the solutions of the new multi-block algoirthm have been found to be within the order of machine accuracy from those of the baseline sequential algorithm. Single-block runs have been performed on all grids with the two multi-block solvers. The multi-block sequential solver showed an overhead of a few percent, up to 5\% on the coarsest grid, and its parallel counterpart showed an additional overhead of about 1\% in average. The number of restarts has been shown to be directly proportional to the one-dimensional grid size and inversely proportional to the stride size, thus is entirely different from the number of iterations in iterative algorithms. Especially, the number of restarts for $\delta s = \infty$ asymptotically approaches the same linear relationship with the number of partitions in one dimension on all grids, yet again verifying its non-iterative nature. Running with a single process, our new algorithm enabled comprehensively improved performance for essentially all tests compared with the baseline sequential algorithm, and generally the maximum speedups (up to 5 on the finest grid) were obtained at a block size of $64$. For the multi-thread algorithm, substantial parallel speedups with all tests were observed, especially at a block size of $32$. In particular, the speedups were up to 50--80 on a 16-core/32-thread computer, and up to 100--220 on a 64-core/256-thread computer, all on the finest grid. The optimal stride size for the multi-block algorithm was around $2 \Delta h$, same as what found in  \cite{yang2017highly}, and the performance was found to be very robust with regard to the stride size. Furthermore, the block size and stride size, as the free parameters in our new algorithm, barely showed any interdependence in both the sequential and parallel runs. The effects of other aspects such as computing platforms and speed functions on the performance have also been discussed. The extensive study in this paper has thoroughly demonstrated the striking efficiency, flexibility, and applicability of the present algorithm. 

Although beyond the scope of this work, it would be fairly straightforward to implement many extensions such as higher-order upwind schemes, unstructured meshes, anisotropic speeds, etc., in the framework of current block-based restarted narrow band approach, because it still strictly follows the original sequential fast marching method. It would also be interesting to conduct a side-by-side comparison of our algorithm and other algorithms in the literature for solving the Eikonal equations with regard to algorithm complexity and performance on shared-memory computers as a future work.


\begin{thebibliography}{10}

\bibitem{BreussCGV2011}
{\sc M.~Breuss, E.~Cristiani, P.~Gwosdek, and O.~Vogel}, {\em An adaptive
  domain-decomposition technique for parallelization of the fast marching
  method}, Applied Mathematics and Computation, 218 (2011), pp.~32--44.

\bibitem{ChaconV2015}
{\sc A.~Chacon and A.~Vladimirsky}, {\em A parallel two-scale method for
  eikonal equations}, SIAM Journal on Scientific Computing, 37 (2015),
  pp.~A156--A180.

\bibitem{DetrixheGM2013}
{\sc M.~Detrixhe, F.~Gibou, and C.~Min}, {\em A parallel fast sweeping method
  for the eikonal equation}, Journal of Computational Physics, 237 (2013),
  pp.~46--55.

\bibitem{Dijkstra59}
{\sc E.~W. Dijkstra}, {\em A note on two problems in connexion with graphs},
  Numerische Mathematik, 1 (1959), pp.~269--271.

\bibitem{Gillberg2014}
{\sc T.~Gillberg, A.~M. Bruaset, {\O}.~Hjelle, and M.~Sourouri}, {\em Parallel
  solutions of static hamilton-jacobi equations for simulations of geological
  folds}, Journal of Mathematics in Industry, 4 (2014), pp.~1--31.

\bibitem{Herrmann03}
{\sc M.~Herrmann}, {\em A domain decomposition parallelization of the fast
  marching method}, in {Annual Research Briefs}, Center for Turbulence
  Research, Stanford University, Stanford, {CA}, 2003, pp.~213--225.

\bibitem{JeongW08}
{\sc W.-K. Jeong and R.~T. Whitaker}, {\em A fast iterative method for
  {E}ikonal equations}, SIAM Journal on Scientific Computing, 30 (2008),
  pp.~2512--2534.

\bibitem{Kim01}
{\sc S.~Kim}, {\em An {O}({N}) level set method for {E}ikonal equations}, SIAM
  Journal on Scientific Computing, 22 (2001), pp.~2178--2193.

\bibitem{RouyT92}
{\sc E.~Rouy and A.~Tourin}, {\em A viscosity solutions approach to
  shape-from-shading}, SIAM Journal on Numerical Analysis, 29 (1992),
  pp.~867--884.

\bibitem{Sethian96}
{\sc J.~A. Sethian}, {\em A fast marching level set method for monotonically
  advancing fronts}, Proceedings of the National Academy of Sciences, 93
  (1996), pp.~1591--1595.

\bibitem{Sethian99a}
{\sc J.~A. Sethian}, {\em Fast marching methods}, SIAM Review, 41 (1999),
  pp.~199--235.

\bibitem{Sethian99b}
{\sc J.~A. Sethian}, {\em Level Set Methods and Fast Marching Methods: Evolving
  Interfaces in Computational Geometry, fluid Mechanics, Computer Vision, and
  Materials Science}, Cambridge University Press, Cambridge, $2^{nd}$~ed.,
  1999.

\bibitem{Tsitsiklis95}
{\sc J.~Tsitsiklis}, {\em Efficient algorithms for globally optimal
  trajectories}, IEEE Transactions on Automatic Control, 40 (1995),
  pp.~1528--1538.

\bibitem{WeberDBBK08}
{\sc O.~Weber, Y.~S. Devir, A.~M. Bronstein, M.~M. Bronstein, and R.~Kimmel},
  {\em Parallel algorithms for approximation of distance maps on parametric
  surfaces}, ACM Transactions on Graphics, 27 (2008), pp.~1--16.

\bibitem{yang2017highly}
{\sc J.~Yang and F.~Stern}, {\em A highly scalable massively parallel fast
  marching method for the eikonal equation}, Journal of Computational Physics,
  332 (2017), pp.~333--362.

\bibitem{YatzivBS06}
{\sc L.~Yatziv, A.~Bartesaghi, and G.~Sapiro}, {\em {O}({N}) implementation of
  the fast marching algorithm}, Journal of Computational Physics, 212 (2006),
  pp.~393--399.

\bibitem{Zhao05}
{\sc H.~Zhao}, {\em A fast sweeping method for {E}ikonal equations},
  Mathematics of Computation, 74 (2005), pp.~603--627.

\bibitem{Zhao07}
{\sc H.~Zhao}, {\em Parallel implementations of the fast sweeping method.},
  Journal of Computational Mathematics, 25 (2007), pp.~421--429.

\end{thebibliography}
\end{document}